\documentclass[%
 reprint,
 amsmath,amssymb,
 aps,
prl,
floatfix,
]{revtex4-2}

\usepackage{placeins}
\usepackage{xcolor}
\usepackage[normalem]{ulem}
\usepackage{comment}
\usepackage{graphicx}
\usepackage{dcolumn}
\usepackage{bm}
\usepackage{bbm}
\usepackage{bbold}
\usepackage{multirow}
\usepackage{hyperref}
\hypersetup{colorlinks=true,breaklinks,linkcolor=blue,urlcolor=blue,citecolor=blue}
\usepackage{braket}
\usepackage{comment}
\usepackage{breakcites}
\usepackage{orcidlink}

\newcommand{\TUVienna}{\affiliation{$^a$Institute of Solid State Physics, TU Wien, 1040 Vienna, Austria}}
\newcommand{\UniWuerzburg}{\affiliation{$^c$Institut für Theoretische Physik und Astrophysik and Würzburg-Dresden Cluster of Excellence ct.qmat, Universität Würzburg, 97074 Würzburg, Germany}}
\newcommand{\UniHamburg}{\affiliation{$^b$ I. Institute for Theoretical Physics, Universität Hamburg, Notkestraße 9–11, 22607 Hamburg, Germany}}
\newcommand{\Laquila}{\affiliation{$^d$Dipartimento di Scienze Fisiche e Chimiche, Università dell’Aquila, Coppito-L’Aquila, Italy}}

\newcommand{\pdg}{{\ensuremath{\phantom{\dagger}}}}

\begin{document}

\title{Effective enhancement of the electron-phonon coupling \\ driven by  nonperturbative electronic density fluctuations}

\author{E.~Moghadas$^{a}$ \orcidlink{0009-0001-9441-2123}}
\author{M.~Reitner$^{a}$ \orcidlink{0000-0002-2529-0847}}
\author{T.~Wehling$^{b}$ \orcidlink{0000-0002-5579-2231}}
\author{G.~Sangiovanni$^{c}$ \orcidlink{0000-0003-2218-2901}}
\author{S.~Ciuchi$^{d}$ \orcidlink{0000-0003-1251-1484}}
\author{A.~Toschi$^{a}$ \orcidlink{0000-0001-5669-3377}}

\TUVienna
\UniHamburg
\UniWuerzburg
\Laquila

\date{\today}

\begin{abstract}
We present a dynamical mean-field study of the nonperturbative electronic mechanisms, which may lead to significant enhancements of the electron-phonon coupling in correlated electron systems. Analyzing the effects of electronic correlations on the lowest-order electron-phonon processes, we show that in the proximity of the Mott metal-to-insulator transition of the doped square lattice Hubbard model, where the isothermal charge response becomes particularly large at small momenta, the coupling of electrons to the lattice is strongly increased. This, in turn, induces significant corrections to both the electronic self-energy and phonon-mediated pairing interaction, indicating the possible onset of a strong interplay between lattice and electronic degrees of freedom even for small values of the bare electron-phonon coupling.

\end{abstract}

\maketitle

\textit{\label{Introduction}Introduction.} 
The scattering processes of electrons and lattice vibrations (electron-phonon coupling) represent an important factor in determining  thermodynamic and spectroscopic properties of condensed matter systems \cite{abrikosov2012methods, rickayzen2013green, mahan2013many}. It is also widely known that these processes play a crucial role in the electronic pairing in conventional superconductors both at ambient \cite{abrikosov2012methods, rickayzen2013green, tinkham2004introduction, degennes1999superconductivity} and under high-pressure \cite{floreslivas2020, eremets2024}. In fact, an estimate of the magnitude of the electron-phonon (el-ph) coupling is often decisive \cite{boeri2008} to classify the superconductivity shown by a given compound as ``conventional'' or ``unconventional'', whereas the latter is typically driven by a purely electronic pairing mechanism.
However, while the algorithmic  progress of DFT-based methods in the last decades has allowed to reach an unprecedented level of precision for the quantitative estimates or the prediction of the el-ph coupling and their effects \cite{giustino2017, floreslivas2020}, the corresponding treatment of this physics in strongly correlated electron systems \cite{grilli1994, pietronero1995, arrigoni2003, koch2004, pietronero2004, roesch2004, sangiovanni2005, sangiovanni2006afm, sangiovanni2006dynamics, seibold2007, roesch2007, gunnarsson2008, capone2010rev, chen2021, wang2021, jiang2022, shen2024, velasco2025} remains highly challenging and is still an active area of research \cite{abramovitch2025, coulter2025}.

In this paper, we address this longstanding problem from a fundamental perspective, (i) by performing a rigorous investigation of the renormalization of the el-ph coupling in the intermediate-to-strong coupling regime and (ii) by evaluating the associated effects on the electronic pairing and self-energy. In doing this, we also aim at clarifying the partially contradicting interpretations and results obtained by different groups in very early studies of this problem \cite{arrigoni2003, koch2004, pietronero2004, seibold2007}, which were limited by the computational effort twenty years ago. 

In order to achieve these goals in the most basic framework,  we have performed dynamical mean-field theory (DMFT) calculations, on both the one- and two-particle level, for the (hole-doped) Hubbard model \cite{Hubbard1963, hubbard_rev} on the square lattice with only nearest-neighbor (n.n.) hopping in the close proximity of its Mott metal-to-insulator transition (MIT). The choice of DMFT allows an explicit treatment of the non-perturbative effects  of the electronic interaction \cite{schäfer2013, schäfer2016, gunnarson2017, chalupa2018, vucicevic2018, springer2020, reitner2020attractive, chalupa2021, pelz2023highly, adler2024}, which -as we will demonstrate- play a pivotal role in the renormalization of the el-ph scattering. Further, the analytical insight recently gained on the momentum-dependent response function \cite{reitner2020attractive, reitner2023nh, kowalski2023thermodynamic} of DMFT makes it possible to clarify the underlying origin of our numerical results, and of the emerging physics.

Finally, we elaborate on the consequences and the limitations of our findings also in relation to their applicability to the physics of correlated oxides, Hund's metals and unconventional superconductors, and outline the possible routes they might suggest for future investigations.

\textit{Momentum-dependent charge response.} In our study we will focus on a typical realization of el-ph scattering, where the lattice vibrations are directly coupled to the electronic density. For this reason, we start by considering the (momentum-dependent) charge-response of the electronic system. This quantity describes, in general, the electronic density fluctuations, whereas in the static limit (i.e. zero transfer frequency, $\omega=0$, and transfer momentum $\mathbf{q}\to 0$), yields the (electronic) isothermal compressibility, $\kappa = \frac{1}{n^2} \, \partial n/\partial\mu = \frac{1}{n^2} \chi_{\mathbf{q\to0}}(\omega=0)$ \cite{Watzenboeck2020, Wilcox1968, krien2017, krien2019, vanloon2015}, $n$ being the electron density of the system and $\mu$ its chemical potential.

\begin{figure*}
    \centering
    \includegraphics[width=\textwidth]{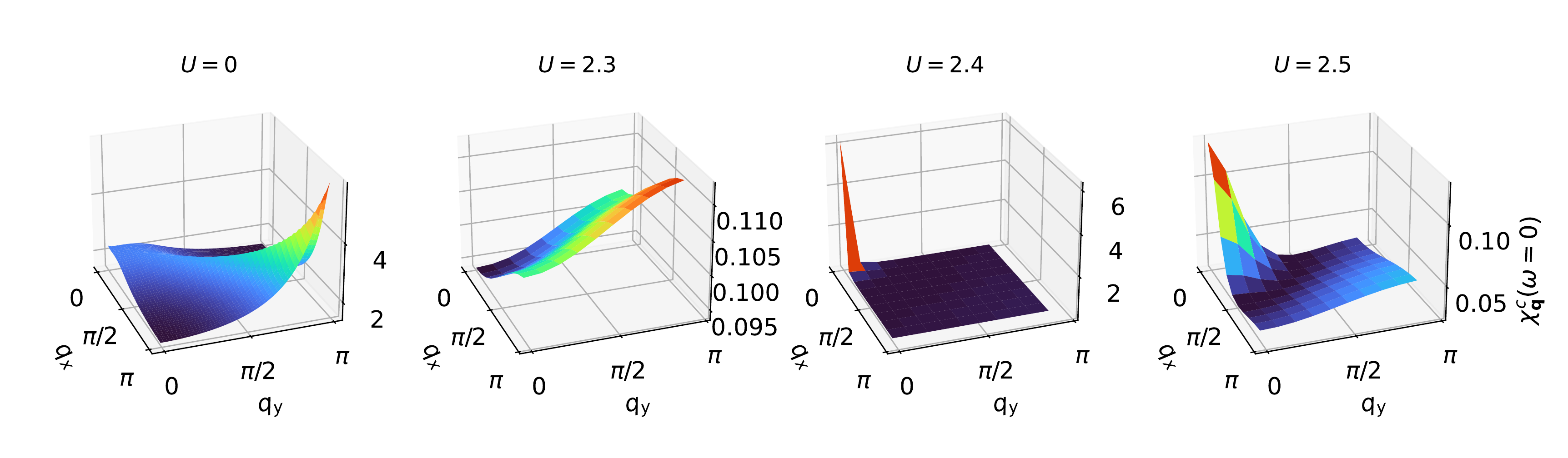}
    \vspace{-10mm}
    \caption{Momentum dependence of the static charge susceptibility $\chi^c_{\mathbf{q}}$ computed in DMFT for the 2D-Hubbard model on the square lattice for $\beta=55.75$ and $n=0.9966$ at the bosonic frequency $\omega=0$ and evaluated for varying $U$ values.} 
    \label{fig:chi_q_U}
\end{figure*}

In this respect, we recall that the pioneering DMFT works of Refs.~\cite{kotliar2002, werner2007, eckstein2007,nourafkan2019} demonstrated that the electronic compressibility may become significantly enhanced in the vicinity of the critical endpoint of the Mott MIT. Specifically, this happens in  slightly doped, as well as frustrated systems \cite{kotliar2002}, where even a divergence of $\kappa$ can be observed. 
At the same time, we should also note how the general understanding of the physical processes controlling this phenomenon has recently advanced by means of the combined analytical/numerical studies in Refs.~\cite{reitner2020attractive,vanloon2020,reitner2023nh,kowalski2023thermodynamic}. In particular, these studies identified the purely non-perturbative mechanisms controlling the significant enhancement of the uniform charge response in parameter regimes, where the on-site charge fluctuations get significantly suppressed by the strong electronic interaction.


Based on these considerations, here we extend the derivation of Refs.~\cite{reitner2020attractive, reitnerDiss, chalupa2022phd, kowalski2023thermodynamic} to general transfer momenta, obtaining 
the following spectral representation, which yields a rather precise description of the lattice charge susceptibility of DMFT for the square lattice in the intermediate-to-strong-coupling regime (see Appendix A in the \textit{End Matter} section):
\begin{equation}\label{equ:kappa_bethe}
   \chi^c_{\mathbf{q}}(\omega \! = \! 0) \equiv \chi^c_{\mathbf{q}} \simeq \sum_{\alpha} \, \left( \lambda_{\alpha}^{-1} + \beta \, {\cal T}_{\mathbf{q}} \right)^{-1} \! w_{\alpha}
\end{equation}
Here, $\lambda_{\alpha}$ denotes the eigenvalues of the generalized {\sl on-site} static charge susceptibility of the system, $\chi_{\rm loc}^{c,\nu\nu'\omega=0}$, expressed as a matrix in the fermionic Matsubara frequency space $\{\nu,\nu'\}$ and the spectral weight $w_{\alpha}$ consists of the associated eigenvectors. The momentum dependence of Eq.~(\ref{equ:kappa_bethe}) is enclosed entirely in the quantity
$ \beta \, \mathcal{T}_{\mathbf{q}} \approx  \beta \,t^2\Bigl[ \cos(q_x) + \cos(q_y) \Bigr] $ (see Appendix A and \cite{supplemental} for more details), with the n.n. hopping amplitude $t$ and the inverse temperature $\beta=1/T$.

While Eq.~\eqref{equ:kappa_bethe} will be quite useful for interpreting the numerical results presented below, as well as a starting point for additional derivations, it immediately allows to  define the condition for a strong enhancement of  $\chi^c_{\mathbf{q}}$, namely when $\lambda^{-1}_\alpha \rightarrow - \beta {\cal T}_{\mathbf{q}}$, for $\omega_{\alpha}\neq0$ \cite{reitner2020attractive}. In particular, according to Eq.~\eqref{equ:kappa_bethe}, the divergence of the electronic compressibility would occur when the lowest eigenvalue of the on-site generalized charge susceptibility $\lambda_I$ becomes negative enough to fulfill the condition $\lambda^{-1}_I \rightarrow - \beta \, {\cal T}_{\mathbf{q}=0} = -2 \beta t^2 < 0$. This is evidently possible only after a sign-flip of $\lambda_I$ (which is typically positive at weak-coupling) occurs, which features a {\sl non-invertibility} of the Bethe-Salpeter equations for the on-site charge susceptibility \cite{schäfer2013, janis2014, schäfer2016, gunnarson2016, chalupa2018, vucicevic2018, thunström2018, springer2020, pelz2023highly, essl2024} and marks a breakdown of the corresponding self-consistent perturbation expansion \cite{kozik2015, gunnarson2017, vucicevic2018, adler2024, essl2025}. As the eigenvalues $\lambda_\alpha$ decrease with the Hubbard interaction $U$, due to the freezing of on-site charge fluctuations induced by the pre-formation of local magnetic moments \cite{gunnarson2016,chalupa2021,mazitov2022localmomentA,mazitov2022localmomentB,adler2024}, Eq.~(\ref{equ:kappa_bethe}) also directly clarifies \cite{reitner2020attractive} how a substantial increase of the ${\bf q}=0$ charge response of the system can be driven, in the non-perturbative regime, by a suppression of local charge fluctuations. 

\textit{Numerical results.} We now turn to our numerical DMFT investigation of the evolution of the charge susceptibility as a function of the interaction $U$. For our DMFT calculations we exploit the continuous time quantum Monte Carlo algorithm in the hybridization expansion as implemented in the \textit{w2dynamics} code-package \cite{w2dyn}. In particular, we consider here the case of the unfrustrated Hubbard model on the square lattice in energy units of its half-bandwidth, i.e. $4t=1$. In order to lift the constraint of perfect particle-hole symmetry, and allow for an enhancement of the compressibility, we specifically choose the filling of $n=0.9966$. This sets us in the close proximity of the critical endpoint of the Mott MIT for the value of $\beta=55.75$ and the range of interactions $U$ considered here. 

In Fig.~\ref{fig:chi_q_U}, we present the full momentum dependence of the charge susceptibility for different values of $U$. We note that, in the non-interacting $U=0$ case (leftmost panel), the charge response is dominated by the contribution of the $\mathbf{q}=(\pi,\pi)$ sector, due to the perfect nesting properties of the square lattice \cite{georges1996} considered. Upon turning the interaction up to $U=2.3$ (second panel), one observes, as generally expected on the basis of perturbation theory arguments (cf.~RPA),  a gradual,  overall suppression of the charge susceptibility for all $\mathbf{q}$-values, while the predominance of larger transfer momenta is still retained. The situation changes drastically once the interaction value of $U=2.4$ (corresponding to a parameter set very close to the half-filled MIT)  is reached (third panel), where a strong enhancement for the isothermal susceptibility occurs: A sharp peak around $\mathbf{q}=(0,0)$ emerges, signaling the tendency toward a phase separation instability. For larger $U$ values (rightmost panel), further proceeding into the bad-metallic region, we continue to observe the peak at low transfer momenta, although with a smaller overall magnitude. The drastic difference in the momentum structure between the metallic ($U<2.4$) and the bad-metallic ($U \geq 2.4$) regime can be quantitatively understood in terms of the spectral representation of the charge response introduced in Eq.~(\ref{equ:kappa_bethe}). Specifically, as explicitly illustrated in \cite{supplemental},  the spectral weight $w_I$, associated to the lowest eigenvalue $\lambda_I$ (responsible for the emergence of the  $\mathbf{q}=(0,0)$ peak \cite{reitner2020attractive}), is vanishingly small in the whole weak-coupling regime, and only obtains a sizable value in the bad-metallic phase, explaining the sudden change in the momentum dependence of $\chi_{\mathbf{q}}^c$ observed at $U=2.4$. 

\textit{Effect on the electron-phonon vertex.}
\begin{figure}
    \centering
    \includegraphics[width=0.48\textwidth]{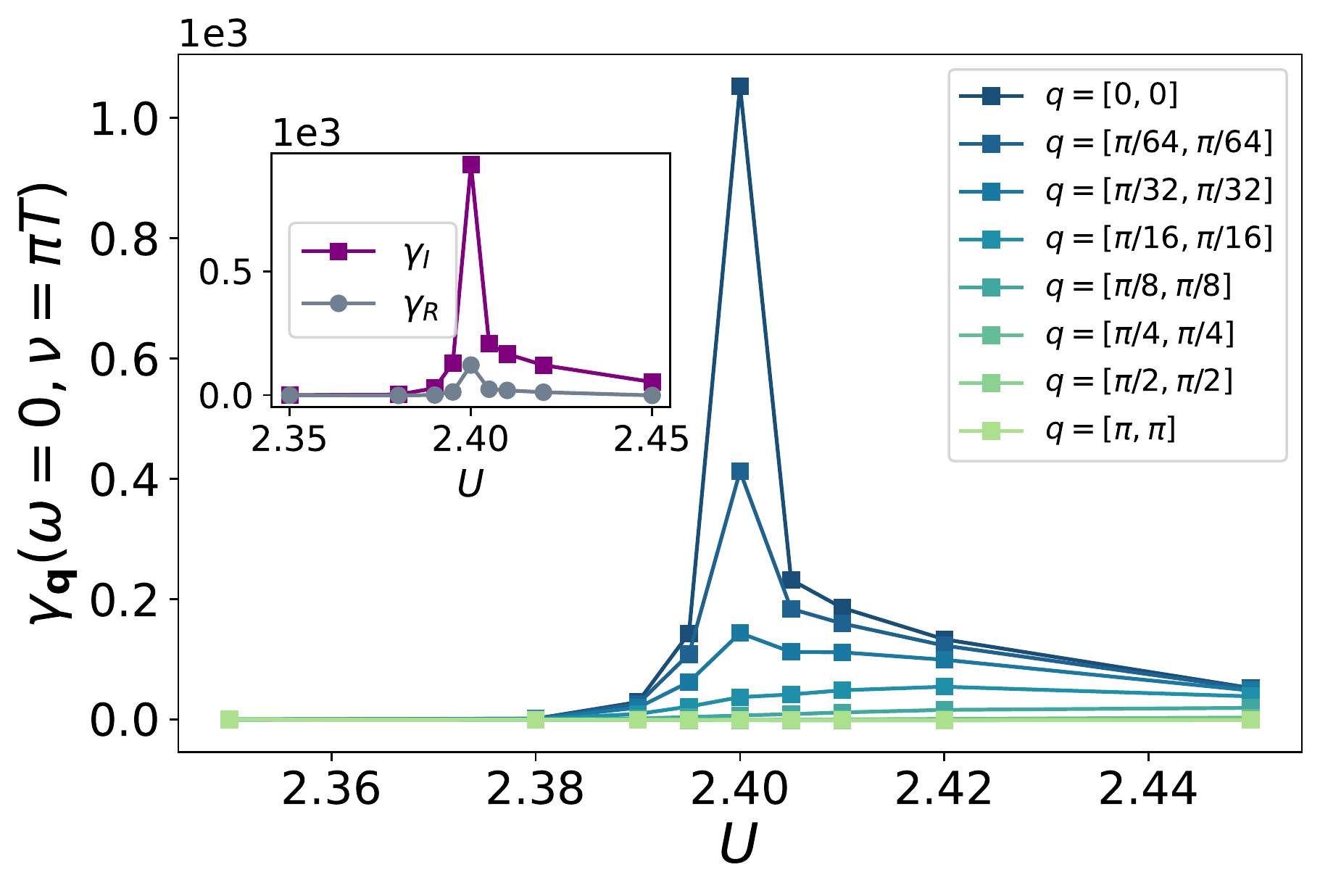}
    \vspace{-8mm}
    \caption{Renormalized el-ph vertex $\gamma_{\mathbf{q}} \!= \! \tilde{g}_{\mathbf{q}}/g_0$ extracted from $\chi^c_{\mathbf{q}}$ as a function of $U$ at the lowest bosonic/fermionic frequencies at $\beta=55.75$. Inset: Contribution of the lowest eigenvalue to $\gamma_{\mathbf{q}}$  ($\gamma_I$), in comparison to the contribution of the remaining eigenvalues ($\gamma_{R}$).}
    \label{fig:gamma_q_U}
\end{figure}
\begin{figure*}
    \centering
    \includegraphics[width=\textwidth]{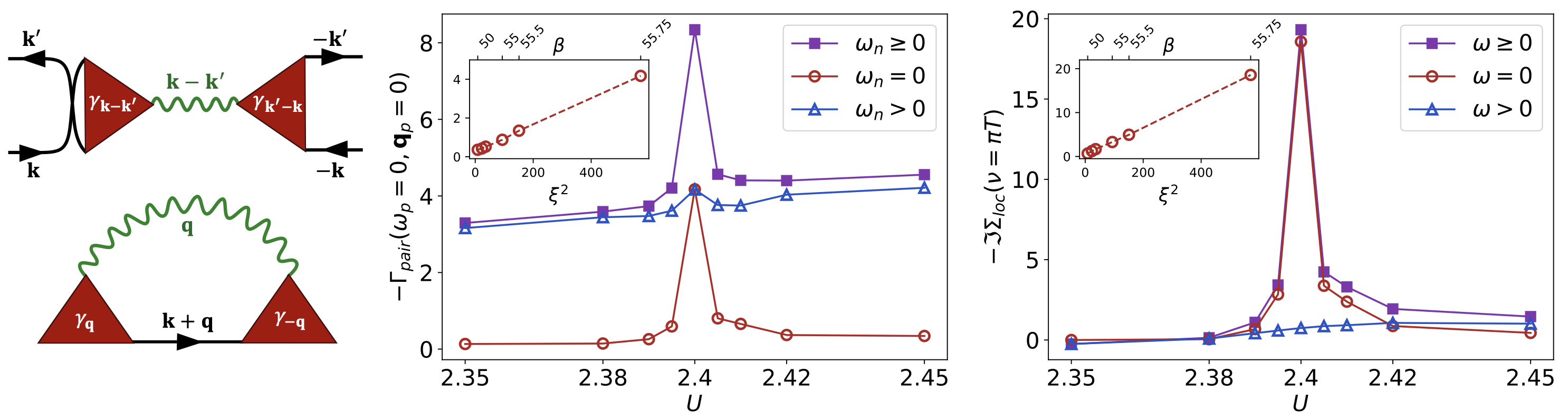}
    \vspace{-8mm}
    \caption{Left panel: Lowest order contributions in $g_0$ to the  phonon-mediated pairing interaction and to the electronic self-energy. Central panel:  $U$-dependence  of the pairing diagram (violet) in units of $\lambda$ for $\beta=55.75$ and $\omega_0 =0.1$, split in its contributions arising from the zero (red) and  finite frequency (blue) components. Inset: Dependence of the zero-frequency contribution to the pairing diagram on the correlation length and inverse temperature. Right panel: Same plots as for the central panel, but for the self-energy corrections (s. text).}
    \label{fig:pairing}
\end{figure*}
After discussing the nonperturbative enhancement of the electronic charge susceptibility, it is natural to pose the question how/to what extent this phenomenon can manifest itself in the renormalization of the el-ph coupling. To this end, we consider the most simple case of optical Holstein phonons with a characteristic frequency of $\omega_0$ and extend the Hubbard Hamiltonian by an additional interaction term with the bare, local coupling $g_0$ defined as $\mathcal{H}_{el-ph}=g_0\sum_{\mathbf{k,q}}c^{\dagger}_{\mathbf{k+q}}c_{\mathbf{k}}(b_{\mathbf{q}}+b_{\mathbf{-q}}^{\dagger})$ where $b \, (b^{\dagger})$ denote the bosonic annihilation (creation) operators. In the spirit of Refs.~\cite{arrigoni2003, koch2004, pietronero2004, seibold2007} we consider purely electronic corrections to the el-ph vertex linear in $g_0$. 
In this work we extend the merely electronic results of Refs.~\cite{reitner2020attractive, reitnerDiss, chalupa2022phd, kowalski2023thermodynamic} to the renormalized el-ph coupling $\tilde{g}$ (cf. Appendix B) and derive, without additional approximations w.r.t Eq.~\eqref{equ:kappa_bethe}, an analytical expression for the dimensionless quantity $\gamma_{\mathbf{q}}^{\nu}:=\tilde{g}_{\mathbf{q}}/g_0$ in terms of the on-site generalized susceptibility's eigenbasis
\begin{equation}\label{equ:gamma_spectral}    \gamma_{\mathbf{q}}^{\nu} \simeq \sum_{\alpha} \left( \lambda_{\alpha}^{-1} + \beta {\cal T}_{\mathbf{q}} \right)^{-1} \tilde{w}^{\nu}_{\alpha} \, .
\end{equation}

The above equation highlights the intrinsically {\sl non-perturbative} connection between the enhancement of the $\mathbf{q} \!= \!0$ charge response and the el-ph coupling. In fact, both quantities share the same divergence condition ($\lambda^{-1}_I \rightarrow - \beta {\cal T}_{\mathbf{q}=0} = - 2\beta t^2 < 0$), while only the spectral weight, $\tilde{w}^{\nu}_{\alpha}$, for the renormalized el-ph vertex gets modified w.r.t the one of the charge response. In fact, as explicitly shown in Appendix C, the proximity to the Mott transition, except for perfect particle-hole symmetry, will always yield a diverging compressibility \emph{and} a strongly enhanced el-ph coupling. These arguments are based on the thermodynamic stability criteria of the Hubbard model independent of dimensionality \cite{kowalski2023thermodynamic, reitnerDiss}. Hence, they do apply but are not restricted to the DMFT data presented here.

Our DMFT results for $\gamma^{\nu}_{\mathbf{q}}$ evaluated at the lowest fermionic frequency $\nu=\pi T$, obtained for the same parameter values as the charge susceptibility, are shown in Fig.~\ref{fig:gamma_q_U}. On the basis of Eq.~(\ref{equ:gamma_spectral}), it is clear that the momentum differentiation between the weakly-correlated/metallic and the bad metallic regimes, previously discussed for the charge susceptibility, gets also directly reflected onto the behavior of $\gamma_{\mathbf{q}}^{\nu}$. Specifically, for $U<2.4$, an extremely reduced renormalized el-ph coupling, which is strongly suppressed $\gamma_{\mathbf{q}}^{\nu= \pi T} \! \ll \! 1$ and essentially independent of the transfer momentum, is found, consistent with the conventional, gradual suppression of the charge response by increasing interactions. However, by further increasing the interaction up to $U\cong 2.4$,  the renormalized el-ph vertex suddenly exhibits a huge enhancement at low transfer momenta $\mathbf{q}$.  This reaches its maximum at $U=2.4$ and then slowly decreases with growing interaction values. 
Consistent with our previous discussions, we observe that this enhancement predominantly originates from the behavior of the lowest, real eigenvalue $\lambda_I$ of the impurity susceptibility (for details cf. \cite{supplemental}), as is shown by the inset of Fig.~\ref{fig:gamma_q_U} where we compare its contribution to the renormalized vertex, $\gamma_I$, with the contribution of the remaining eigenvalues $\gamma_R$. 
At the same time, while the behavior of $\gamma_{\mathbf{q}}^{\nu= \pi T}$ in this regime is evidently controlled by the underlying evolution of 
$\chi_{\mathbf{q}}^c$, we observe nonetheless a relative enhancement of the renormalized el-ph coupling (w.r.t.~the non-interacting case, where $\gamma_{\mathbf{q}}^{\nu= \pi T} \equiv 1$) which is larger by {\sl two orders} of magnitude than for the corresponding charge response. This additional enhancement can be ascribed to the prefactor proportional to the inverse dressed DMFT bubble term $\chi_{0,\mathbf{q}}^{\nu}$ in Eq.~\eqref{equ:eph_vertex_app} of Appendix B, as it gets strongly suppressed at low frequencies in the bad-metallic regime. This extra factor, required to obtain the renormalized coupling from the charge susceptibility, might have caused some contradicting interpretations in the past literature, since in the region of small but finite {\bf q}, the enhancement of $\gamma_{\mathbf{q}}^{\nu= \pi T} $ can become numerically visible even {\sl before} the corresponding increase of $\chi_{\mathbf{q}}^c$ \cite{supplemental}.

In any case, the overall similarity between $\gamma_{\mathbf{q}}^{\nu}$ and $\chi_{\mathbf{q}}^c$ in the proximity of the Mott MIT suggests that the former may display a momentum-dependence characteristic for an Ornstein-Zernike (OZ) function, which represents, in fact, a reliable parametrization for the isothermal charge compressibility computed in DMFT in the vicinity of its second-order phase transition \footnote{It is important to emphasize here, that, different from, e.g., antiferromagnetic phase-transitions, the phase-separation instability relevant for our study belongs to the Ising universality class. Hence, it will not be necessarily destroyed by spatial fluctuations beyond DMFT even in the 2D system considered.}. As solely $\mathcal{T}_{\mathbf{q}}$ is responsible for the momentum structure of Eq.~\eqref{equ:kappa_bethe} and Eq.~\eqref{equ:gamma_spectral}, its expansion for small transfer momenta, ${\cal T}_{\mathbf{q}} \approx t^2(2 - \frac{1}{2}  \mathbf{q}^2) + \mathcal{O}(\mathbf{q}^4)$ allows to derive an explicit expression for the correlation length of the charge response in terms of the eigenvalues of the local impurity susceptibility \cite{supplemental}
\begin{equation}\label{equ:xi_chi_ev}
    \xi^2 \simeq \frac{\sum_{\alpha} \xi_{\alpha}^2\chi_{\alpha}w_{\alpha}}{\sum_{\alpha}\chi_{\alpha}w_{\alpha}}.
\end{equation}
where $\chi_{\alpha} = \Bigl( \lambda_{\alpha}^{-1} + \beta \mathcal{T}_{\mathbf{q}=0} \Bigr )^{-1}$ and $\xi^2_{\alpha} = -\frac{\beta t^2}{2} \,\chi_{\alpha}$. For the renormalized el-ph vertex we merely need to substitute $w_{\alpha}\to\tilde{w}_{\alpha}$ in Eq.~\eqref{equ:xi_chi_ev}. It is worth noting, that in a first approximation, close to the MIT, where the physical behaviors are dominated by the contribution of the lowest eigenvalue $\lambda_I, $ the correlation lengths of $\gamma_{\mathbf{q}}^{\nu= \pi T}$ and $\chi_{\mathbf{q}}^c$ will become equivalent \cite{supplemental}.

\textit{Renormalization of the electronic quantities.}
After addressing the effect of enhanced charge fluctuations on the el-ph coupling, we turn to analyze the role this strong renormalization plays on the electronic system. To this end, we start by considering the effective electron-electron pairing interaction mediated by the exchange of one single phonon taking the aforementioned vertex renormalization effects into account. 
The corresponding pairing diagram, depicted in the upper left panel of Fig.~\ref{fig:pairing}, can be thus viewed as a {\sl perturbative} correction to the electronic scattering in the pairing  up to {\sl second order} in the bare el-ph coupling. This may provide a reasonable starting point for analyzing the regime of small bare el-ph coupling in the presence of strong electronic correlations ($g_0 \ll U$).

For zero transfer energy and momentum in the pairing channel (i.e. $\omega_p=0, \mathbf{q}_p = 0$) and after a change of variables in the frequency and momentum arguments this interaction can be written as
\begin{align}\label{equ:gamma_integral}
    & \Gamma_{pair} = \frac{g_0^2}{\beta^2} \sum_{\nu\omega}\int d^2q \, \gamma_{\mathbf{-q}}^{-\nu, -\omega} \, D_0^\omega \, \gamma_{\mathbf{q}}^{\nu\omega} \\
    \label{equ:gamma_integral2}
    &{}\cong \frac{-2g_0^2}{\omega_0\,\beta^2}\sum_{\nu} \int d^2q \frac{\gamma_{\mathbf{q}=0}^{\nu} \gamma_{\mathbf{q}=0}^{-\nu}}{\bigl[1+(\xi\mathbf{q})^2\bigr]^2} + \sum_{\omega>0} \mathcal{R}(\omega)
\end{align}
where in the last line we assumed an OZ parametrization for the el-ph vertex, i.e. $\gamma_{\mathbf{q}} \propto \xi^2\,\bigl[1+(\xi\mathbf{q})^2 \bigr]^{-1}$ \footnote{Rigorously speaking, one would expect the OZ momentum dependence of $[\chi_{\mathbf{q}}^c]^{-1}$ or ($[\gamma_{\mathbf{q}}]^{-1}$) to be corrected as $|{\bf q}|^{2-\eta}$, $\eta = \frac 14$ being the corresponding anomalous exponent of the 2D Ising class \cite{huang1987}.}. Here, $D_0^{\omega} = \frac{-2\,\omega_0}{\omega_n^2+\omega_0^2}$ denotes the bare phonon propagator (with $\omega_0=0.1$) and $\mathcal{R}(\omega)$ is a regular function of the positive bosonic Matsubara frequencies.
A quick glance at the momentum integral immediately reveals an infrared singularity for $\xi\to\infty$. This yields a divergence of $\Gamma_{pair}$ that, in 2D, scales with the square of the correlation length $\xi$ \footnote{Note that even the explicit inclusion of the anomalous exponent $\eta$ in the OZ expression would not prevent such a divergence, which would occur with a slightly modified scaling, $\sim \xi^{\frac 32}$}. 
In the central panel of Fig.~\ref{fig:pairing}, we show then our numerical data for $\Gamma_{pair}$ in units of the effective el-ph coupling strength $\lambda=\frac{2g_0^2}{\omega_0}$, evaluated using Eq.~(\ref{equ:gamma_integral}) as a function of $U$ and decompose it into the contributions arising from the static ($\omega \! = \! 0$) and dynamic ($\omega \! > \! 0 $) components. We clearly observe the effect of the static contribution to the overall behavior of $\Gamma_{pair}$, namely a prominent peak at the critical interaction value of $U=2.4$ in contrast to a rather weak-$U$ dependence of the ($\omega \! > \! 0 $) terms \footnote{This difference is due to the purely static nature of $\chi^c
_{\mathbf q =0}(\omega)$, which directly reflects the conservation of the total electric charge in the system.}. Moreover, we can analyze the dependence of $\Gamma_{pair}$ on the correlation length by evaluating its (strongly enhanced) static contribution for different values of $\beta$ (s.~inset of Fig.~\ref{fig:pairing}, central panel). As expected from the (approximated) Eq.~\eqref{equ:gamma_integral2}, the numerical evaluation of this quantity reveals a $\xi^2$-dependence.

Among the several Feynman diagrams, entailing second-order corrections in $g_0$ to the DMFT (purely electronic) self-energy, which display the leading contribution in $\xi$ in the proximity of the phase-separation instability \footnote{A more detailed discussion of the different class of el-ph diagrams, and their scaling behavior in $\xi$, based on a rather general treatment à la ab-initio D$\Gamma$A \cite{toschi2011} can be found in \cite{supplemental}}, we focus on the one depicted in the lower left panel of Fig.~\ref{fig:pairing}, which was often considered in previous works \cite{roesch2007, gunnarsson2008, rubtsov2012, vanloon2014, vandelli2022} and which displays an evident diagrammatic similarity to the pairing interaction process discussed above.
By following essentially the same step as for the calculations of $\Gamma_{pair}$, we get:
\begin{align}
 \label{equ:self1}
    &\Delta \Sigma_{\mathbf{k}}^{\nu} = \frac{g_0^2}{\beta} \sum_{\omega}\int d^2q \, \gamma_{\mathbf{-q}}^{\nu+\omega, -\omega} \, G_{\mathbf{k+q}}^{\nu+\omega} \, D_0^\omega \, \gamma_{\mathbf{q}}^{\nu\omega} \\
     \label{equ:self2}
    &{}\cong \frac{-2g_0^2}{\omega_0\,\beta} \int d^2q \frac{G_{\mathbf{k+q}}^{\nu} \,(\gamma_{\mathbf{q}=0}^{\nu})^2}{\bigl[1+(\xi\mathbf{q})^2\bigr]^2} + \sum_{\omega>0} \mathcal{R}(\omega)
\end{align}
Due to the weak $\mathbf{k}$-dependence of the Green's function in the strongly correlated regime, the 2D integral will result, similarly as for $\Gamma_{pair}$, in a $\xi^2$-dependence. These considerations are confirmed by our numerical evaluation of Eq.~(\ref{equ:self1}) reported in the right panel of Fig.~\ref{fig:pairing}. In particular, for the sake of conciseness, we plot the local corrections to the DMFT self-energy  (i.e. $\int d^2k\,\Delta \Sigma_{\mathbf{k}}^{\nu}$)  as a function of $U$ as well as, in the inset, its static contribution as a function of $\beta$. The former data set indicates a huge enhancement of the el-ph corrections to the DMFT self-energy, while the latter confirms the expected $\xi^2$ scaling in the proximity of the non-perturbative phase-separation instability of the electronic system. 

\textit{Conclusions.}
Our analysis based on DMFT calculations and exact thermodynamic considerations demonstrates that a large renormalization of an initially weak bare el–ph coupling can be triggered by enhanced small-$\mathbf{q}$ charge fluctuations in the proximity of a Mott metal–insulator transition. This behavior contrasts the commonly expected suppression of the el–ph coupling arising from reduced charge fluctuations and the suppression of spectral weight at the Fermi surface.
The intrinsically strong-coupling, non-perturbative mechanism responsible for the enhancement of the el–ph scattering identified here is different from that active in (K-doped) BaBiO$_3$, BaSbO$_3$, and BaPbO$_3$, where the enhancement originates instead from the long-range character of the unscreened interaction, which can be captured by \textit{GW} and screened-hybrid-functional DFT approaches \cite{yin2013, li2019, yuan2022}.

The enhancement of the el–ph coupling is directly reflected in a corresponding increase of the effective pairing interaction. For the single-orbital case—possibly relevant to the low-energy physics of cuprate \cite{müller1986, anderson87, rice88, scalapino2012} and nickelate superconductors \cite{hwang2019nickelates, held2020}—this effect is confined to a narrow region of parameter space. The situation may, however, become substantially more favorable in multiorbital systems \cite{grilli1991, grilli1994, mahony2022, labreuil2025}. In particular, we anticipate significant consequences for Hund's metals \cite{haule2009, demedici2011janus, mandal2014, demedici2017hund, arribi2018, isidori2019, Watzenboeck2020}, where the interplay between Hund's and Hubbard interactions favors the formation of large magnetic moments in a strongly correlated, yet resilient metallic phase. In this situation, DMFT-based calculations \cite{chatzieleftheriou2023, georges2013, kowalski2023thermodynamic} show a tendency toward phase-separation instabilities in extended regions of the phase diagrams.

At the same time, the comparably large magnitude of the self-energy corrections estimated within our approach indicates that, for sufficiently large $\xi$, the second-order diagrammatic treatment adopted in this work becomes inadequate. This calls for future self-consistent investigations of the Hubbard–Holstein problem \cite{berger95, sangiovanni2005, sangiovanni2006afm, sangiovanni2006dynamics, macridin2006, capone2010rev, costa2020, abramovitch2025, coulter2025}, for instance based on diagrammatic extensions of DMFT \cite{rohringer2018}.

\begin{acknowledgments}
\textit{Acknowledgements.} 
We thank Aiman Al Eryani, Sabine Andergassen, Massimo Capone, Emmanuele Cappelluti, Sergio Caprara, Herbert Eßl, Samuele Giuli, Alexander Kowalski, Friedrich Krien and Evgeny Stepanov for useful discussions.
E.M., M.R. and A.T. acknowledge support from the Austrian Science Fund (FWF) through the grant 10.55776/I5868 (Project P1 of the research unit QUAST, for5249, of the German Research Foundation, DFG). M.R. further acknowledges support as a recipient of a DOC fellowship of the Austrian Academy of Sciences. G.S. and T.W. acknowledge support from the DFG through FOR 5249-449872909 (Project P5). S.C. acknowledges funding from  NextGenerationEU National Innovation Ecosystem grant ECS00000041 - VITALITY - CUP E13C22001060006 and grant PE00000023 - IEXSMA - CUP E63C22002180006. At this point we also want to thank the WWAQ-organization for their hospitality. Calculations have been performed on the Austrian Scientific Cluster (ASC). 
\end{acknowledgments}


\bibliography{bib/MA_lib, bib/library} 

\begin{thebibliography}{115}
\bibitem [{dat()}]{data-repo}%
  \BibitemOpen
  \href@noop {} {\bibinfo  {journal} {Data set will be made available upon publication of the manuscript}\ }\BibitemShut {NoStop}%
\end{thebibliography}


\clearpage 
\onecolumngrid 
\setcounter{page}{1}
\setcounter{secnumdepth}{1}

\begin{center}
    \textbf{\large End Matter}\\[0.5em]
\end{center}

\thispagestyle{empty}

\twocolumngrid

\textit{Appendix A: Spectral representation of the generalized susceptibility in DMFT - }
The calculation of Bethe-Salpeter equations (BSE) in DMFT is based on the assumption of locality of its kernel, i.e. the irreducible vertex function $\Gamma$ \cite{georges1996, delre2021}. 
For instance, in the charge sector of our interest, one can compute the corresponding static ($\omega=0$) generalized  susceptibility $\chi_{loc}^{c, \nu \nu'}$ of the self-consistently determined AIM \cite{georges1996}, and readily extract the associated $\Gamma_c^{\nu\nu'}$ via the (inverse) BSE \cite{rohringer2012, rohringer2014phd, rohringer2018}
\begin{equation}\label{equ:gamma_loc}
    \frac{1}{\beta^2}\Gamma_c^{\nu \nu'} = [\chi_{loc}^{c, \nu \nu'}]^{-1} - [\chi_{loc}^{0, \nu\nu'}]^{-1} ,
\end{equation}
where the local bubble term is defined as $\chi^{\nu \nu'}_0 = -2\beta G(\nu) \,G(\nu) \, \delta_{\nu \nu'}$, $G(\nu)$ being the local one-particle Green's function. We recall here that the physical impurity charge response can be obtained from the generalized quantity via
\begin{equation}\label{equ:gen_susc_app}
    \chi^{c}_{loc} = \frac{1}{\beta^2}\sum_{\nu\nu'}\chi^{c,\nu\nu'}_{loc} = \sum_{\alpha}\lambda_{\alpha} w_{\alpha} \, ,
\end{equation}
where in the last equality we have decomposed the susceptibility in its spectral representation with eigenvalues $\lambda_{\alpha}$ and eigenvectors $V_{\alpha}^{\nu}$ constituting the spectral weights $w_{\alpha}=\Bigl[ \sum_{\nu} V^{\nu}_{\alpha}\Bigr]\Bigl[ \sum_{\nu'} V^{\nu',\,-1}_{\alpha}\Bigr]$ \cite{gunnarson2017, springer2020}. 

Once having determined $\Gamma_c$, the physical charge susceptibility of the lattice can be obtained from the momentum dependent BSE
\begin{equation}\label{equ:bse_q}
    \chi_{\mathbf{q}}^{c} = \frac{1}{\beta^2}\sum_{\nu \nu'}\Bigl\{ [\chi_{0,\mathbf{q}}^{\nu\nu'}]^{-1} + \frac{1}{\beta^2}\Gamma_c^{\nu\nu'} \Bigr\}^{-1}
\end{equation}
where the momentum dependent bubble susceptibility is defined as $\chi^{\nu\nu'}_{0,\mathbf{q}} = -2\beta\sum_{\mathbf{k}}G_{\mathbf{k}}^{\nu}G_{\mathbf{k+q}}^{\nu}\delta_{\nu\nu'}$ with $G_{\mathbf{k}}^{\nu}$ being the dressed DMFT Green's function. 
Plugging Eq.~\eqref{equ:gamma_loc} into Eq.~\eqref{equ:bse_q}, one obtains the following expression \cite{georges1996}
\begin{equation}\label{equ:bse_dmft}
    \begin{aligned}
        {}& \chi^{c}_{\mathbf{q}} \\
        {}& = \frac{1}{\beta^2} \sum_{\nu \nu'} \left\{[\chi_{loc}^{c, \nu \nu'}]^{-1} + \bigl([\chi^{\nu\nu'}_{0,\mathbf{q}}]^{-1} - [\chi_{0,loc}^{\nu\nu'}]^{-1} \bigr) \right\}^{-1} .
    \end{aligned}
\end{equation}
In the notable case of the infinite-dimensional Bethe lattice the expression for the inverse bubble differences for the uniform and the staggered ordering vector $\mathbf{Q}=(0,0,0,...)$ and $\mathbf{Q}=(\pi, \pi, \pi,...)$ can be expressed analytically \cite{georges1996, reitner2020attractive}
\begin{equation}\label{equ:bub_dif_bethe_pm}
[\chi^{0, \nu \nu'}_{\mathbf{Q}}]^{-1} - [\chi_{loc}^{0, \nu \nu'}]^{-1} = \pm \frac{t^2}{2\beta} \,\delta_{\nu \nu'}      \, ,
\end{equation}
where the plus and minus sign are associated to the uniform and staggered response functions respectively. Since this expression is real valued and constant in the fermionic frequency space, the BSE of Eq.~\eqref{equ:bse_dmft} (for $\mathbf{q}=0$) can be recast in the eigenbasis of the local charge susceptibility defined in Eq.~\eqref{equ:gen_susc_app} \cite{reitner2020attractive, reitnerDiss, chalupa2022phd}
\begin{equation}\label{equ:kappa_bethe_exact}
   \chi^c_{\mathbf{q}=0} = \sum_{\alpha} \, \left( \lambda_{\alpha}^{-1} + \frac{\beta\,t^2}{2} \right)^{-1} \! w_{\alpha}.
\end{equation}
Note that $\chi_{\mathbf{q}=0}^c$ can diverge if the smallest eigenvalue $\lambda_{I}^{-1}\to-\beta t^2/2$ unless for symmetry reasons $w_I$ vanishes (e.g. for perfect particle-hole ($ph$) symmetry at half-filling).

In the case of the 2D square lattice studied here Eq.~\eqref{equ:bub_dif_bethe_pm} can be used, however, as an approximation by replacing $t^2$ with an effective $\mathcal{T}_{\mathbf{q}}$ term:
\begin{equation}\label{equ:teff_bub_app}
\beta^2 \left( [\chi^{0, \nu \nu'}_{\mathbf{q}}]^{-1} - [\chi_{loc}^{0, \nu \nu'}]^{-1} \right) = \beta \mathcal{T}_{\mathbf{q}}(\nu) \delta_{\nu\nu'} \approx  \beta \mathcal{T}_{\mathbf{q}}\delta_{\nu\nu'}.
\end{equation}
where now, in contrast to Refs.~\cite{reitner2020attractive, reitnerDA, reitnerDiss, chalupa2022phd, kowalski2023thermodynamic}, we extend the derivation to general transfer momenta $\mathbf{q}$. Using the above approximation in Eq.~\eqref{equ:kappa_bethe_exact} we arrive at Eq.~\eqref{equ:kappa_bethe} of the main text. 

Importantly, the relatively large selfenergy in the strong coupling limit allows a high frequency expansion of Eq.~\eqref{equ:teff_bub_app} \cite{supplemental} resulting in the following compact expression for $\mathcal{T}_{\mathbf{q}}$
\begin{equation}
        \mathcal{T}_{\mathbf{q}}\approx t^2\Bigl[ \cos(q_x) + \cos(q_y) \Bigr].
\end{equation}
The above equation quantitatively captures how the sign flip of $\mathcal{T}_{\mathbf{q}}$ from $\mathbf{q}=(0,0,...)$ to $\mathbf{q}=(\pi,\pi,...)$ occurs. Additionally, as it yields a real and frequency-independent function, it justifies employing Eq.~\eqref{equ:teff_bub_app} in the strong coupling regime for the quantitative analyses of the main text. 

\textit{Appendix B: Spectral representation of the renormalized electron-phonon coupling in DMFT - }
The renormalized electron-phonon vertex function is defined as the ratio between the renormalized and bare coupling \cite{supplemental}
\begin{equation}\label{equ:eph_vertex_app}
     \gamma_{\mathbf{q}}^{\nu} = \tilde{g}^{\nu}_{\mathbf{q}}/g_0 =\Bigl[ \chi_{0,\mathbf{q}}^{\nu} \Bigr]^{-1} \, \sum_{\nu'} \chi_{\mathbf{q}}^{c,\nu\nu'}.
\end{equation}
We show here that, similarly to the lattice response function, the quantity of interest in our work, $\gamma_{\mathbf{q}}$, can be also expressed in the eigenbasis of the local susceptibility within the DMFT approach. Starting from the generalized momentum dependent susceptibility in this eigenbasis 
\begin{equation}\label{equ:gen_chi_eigen}
    \chi^{c,\nu\nu'}_{\mathbf{q}} \cong \beta^2 \, \sum_{\alpha} V_{\alpha}(\nu) \left( \lambda_{\alpha}^{-1} + \beta \mathcal{T}_{\mathbf{q}} \right)^{-1} V^{-1}_{\alpha}(\nu')
\end{equation}
one can obtain the corresponding expression for the renormalized coupling by inserting Eq.~\eqref{equ:gen_chi_eigen} in Eq.~\eqref{equ:eph_vertex_app}
\begin{equation}\label{equ:eph_lambda_app}
    \begin{aligned}
        {}& \gamma_{\mathbf{q}}^{\nu} \cong \beta^2 \sum_{\alpha, \nu'} (\chi^{\nu}_{0,\mathbf{q}})^{-1} V_{\alpha}(\nu) \left( \lambda_{\alpha}^{-1} + \beta \mathcal{T}_{\mathbf{q}} \right)^{-1} V_{\alpha}^{-1}(\nu') \\
        {}&= \sum_{\alpha} \left( \lambda_{\alpha}^{-1} + \beta {\cal T}_{\mathbf{q}} \right)^{-1} \tilde{w}^{\nu}_{\alpha}
    \end{aligned}
\end{equation}
with  $\tilde{w}^{\nu}_{\alpha} = \beta^2(\chi^{\nu}_{0,\mathbf{q}})^{-1} V_{\alpha}(\nu) \sum_{\nu'}V_{\alpha}^{-1}(\nu')$ denoting the modified spectral weight.

\textit{Appendix C: Thermodynamic stability and its general consequences - }
The thermodynamic stability for the Hubbard model with interaction $U$ is determined by the following conditions ~\cite{kowalski2023thermodynamic, reitnerDiss}
\begin{equation}\label{equ:td_cond}
    \frac{\partial n}{\partial \mu} > 0, \quad -\frac{\partial n}{\partial \mu}\,\frac{\partial D}{\partial U} - \Bigl[ \frac{\partial n}{\partial U} \Bigr]^2 > 0.
\end{equation}
where $n$ is the total density of the system and $D=\braket{n_{\uparrow}n_{\downarrow}}$ denotes the double occupation. Of particular interest is the static ($\omega=0$) charge response of the system, which can be obtained from the corresponding generalized one $\chi_{kk'\mathbf{q}}^c$ [$k=(\nu,\mathbf{k})$] via
\begin{equation}\label{equ:gen_susc_td}
    \chi_{\mathbf{q}}^c = \frac{1}{\beta^2}\sum_{kk'}\chi_{kk'\mathbf{q}}^c = \sum_{\alpha}\lambda^{\alpha}_{\mathbf{q}} w^{\alpha}_{\mathbf{q}}
\end{equation}
where the last equality, differently from Appendix A, now represents the eigendecomposition of the \textit{exact} lattice response function with (momentum-dependent) eigenvalues $\lambda_{\mathbf{q}}^{\alpha}$ and eigenvectors $V_{k\mathbf{q}}^{\alpha}$ associated to the corresponding spectral weights $w^{\alpha}_{\mathbf{q}}=\Bigl[ \frac{1}{\beta}\sum_k V_{k\mathbf{q}}^{\alpha}\Bigr]\Bigl[ \frac{1}{\beta}\sum_{k'} V_{k'\mathbf{q}}^{\alpha-1}\Bigr]$. While the isothermal compressibility is directly related to the physical charge susceptibility, $\partial n/ \partial \mu = 2\,\chi_{\mathbf{q}}^c$, the other thermodynamic derivatives of Eq.~\eqref{equ:td_cond}, $dn/dU:= \partial n / \partial U \bigr|_{\mu}$ and $dD/dU:= \partial D / \partial U \bigr|_{\mu}$, can be obtained from \cite{kowalski2023thermodynamic, reitnerDiss}
\begin{equation}\label{equ:dndu}
    \frac{d n}{d U} = \frac{1}{\beta} \sum_{k\sigma}\frac{dG_{k\sigma}}{dU}
\end{equation}
\begin{equation}\label{equ:dddu}
    \frac{d D}{d U} = -\frac{D}{U} + \frac{1}{2U} \frac{1}{\beta} \sum_{k\sigma}\bigl[ i\nu+\mu-\epsilon_{\mathbf{k}}\bigr]\frac{dG_{k\sigma}}{dU}
\end{equation}
where the derivative of the Green's function with respect to the Hubbard interaction 
can be directly related to the generalized charge susceptibility \cite{kowalski2023thermodynamic, reitnerDiss}
\begin{equation}\label{equ:dg_du}
    \sum_{\sigma}\frac{d G_{k\sigma}}{d U} = \frac{1}{\beta}\sum_{k'} \chi^c_{kk'\mathbf{q}=0}\Bigl( - \frac{\partial \Sigma_{k'}}{\partial U} \Bigr|_G\Bigr)
\end{equation}

 The above equations precisely highlight how the Mott transition, where the double occupancy $D$ becomes discontinuous as a function of $U$, with a divergence of $ \frac{d D}{d U}$ at the critical endpoint of the transition, is fundamentally related to the static, uniform charge response of the system. In fact, the only singular contribution in Eqs.~\eqref{equ:dndu} and ~\eqref{equ:dddu} originates from Eq.~\eqref{equ:dg_du}. In turn, at finite $T$, such a singularity can be only driven by a divergent eigenvalue $\lambda^{\alpha}_{\mathbf{q}=0}$ of $\chi^c_{kk'\mathbf{q}=0}$, and hence, except for perfect $ph$ symmetry, by a corresponding divergence of $\chi^c_{\mathbf{q}=0}$ (cf.~Eq.~\eqref{equ:gen_susc_td}).


Here, we will derive the intrinsic relation linking the Mott transition and the el-ph coupling renormalization. For this,  we consider a general bare coupling $g_0^{kq}$, for which its renormalized counterpart $g^{kq}$ can be extracted from the generalized susceptibility of Eq.~\eqref{equ:gen_susc_td}
\begin{equation}\label{equ:eph_td}
    \begin{aligned}
        g^{kq} &= \Bigl(\chi_0^{kq} \Bigr)^{-1}\sum_{k'}\,\chi^c_{kk'q}\,g_0^{k'q} \\
        &{}=\sum_{\alpha}\lambda_{q}^{\alpha} \, \Bigl(\chi_0^{kq} \Bigr)^{-1} V_{kq}^{\alpha}\sum_{k'}\,V_{k'q}^{\alpha,-1}\,g_0^{k'q} := \sum_{\alpha}g_{\alpha}^{kq}
    \end{aligned}
\end{equation}
Defining $V_{k'q}^{\alpha,-1}\,g_0^{k'q}:=\tilde{V}_{k'q}^{\alpha}$, $- \frac{\partial \Sigma_{k'}}{\partial U} \Bigr|_G:=S_{k'}$ and by inserting $\mathbb{1} = \chi_0^{k\mathbf{q}=0}\Bigl( \chi_0^{k\mathbf{q}=0} \Bigr)^{-1}$ and $\mathbb{1} = \sum_{k''}\tilde{V}_{k''\mathbf{q}=0}^{\alpha}\Bigl( \sum_{k''} \tilde{V}_{k''\mathbf{q}=0}^{\alpha} \Bigr)^{-1}$ we can rewrite Eq.~\eqref{equ:dg_du} in terms of the renormalized coupling of Eq.~\eqref{equ:eph_td}
\begin{equation}\label{equ:dg_du_eph}
    \sum_{\sigma}\frac{d G_{k\sigma}}{d U} = \frac{1}{\beta}\sum_{\alpha}\chi_0^{k\mathbf{q}=0} \, g_{\alpha}^{k\mathbf{q}=0} \, \tilde{\sigma}_{\alpha}^{\mathbf{q}=0}
\end{equation}
where $\tilde{\sigma}_{\alpha}^{\mathbf{q}=0} = \Bigl( \sum_{k''} \tilde{V}_{k''\mathbf{q}=0}^{\alpha} \Bigr)^{-1}\sum_{k'}\tilde{V}_{k'\mathbf{q}=0}^{\alpha}\,S_{k'}$. Here, we assume that particle-hole symmetry or other special symmetry constraints do not apply such that $\sum_{k''} \tilde{V}_{k''\mathbf{q}=0}^{\alpha}\neq0$.

The extension of this important result to the more general case of a  multi-orbital Hubbard Hamlitonian is straightforward. To that end, we will now consider a general, multiorbital electron-phonon coupling $g_{0,a b}^{k q}$, renormalized via
\begin{equation}\label{equ:ren_eph_mo}
    \begin{aligned}
        g_{ab}^{kq} &= \sum_{k', cdef} \Bigl( \chi_{0, cdba}^{kq}\Bigr)^{-1} \, \chi_{cdef}^{c,kk'q} \, g_{0,ef}^{kq} \\
        &{}= \sum_{\alpha} \lambda_q^{\alpha} \sum_{k', cdef} \Bigl( \chi_{0, cdba}^{kq}\Bigr)^{-1} V_{kq, cd}^{\alpha} \, \tilde{V}_{k'q, ef}^{\alpha} := \sum_{\alpha} \, g_{\alpha, ab}^{kq},
    \end{aligned}
\end{equation}
with $\tilde{V}_{k'q, ef}^{\alpha} = V_{k'q, ef}^{\alpha, -1}\,g_{0,fe}^{k'q}$. Following the same steps as for Eq.~\eqref{equ:dg_du_eph}, we obtain
\begin{equation}\label{equ:dg_du_eph_mo}
    \begin{aligned}
        \frac{dG_{k, ab}}{dU} &= \frac{1}{\beta} \sum_{k'cd} \chi_{abcd}^{c,kk'\mathbf{q}=0} \, \Bigl( \frac{-\partial \Sigma_{k',cd}}{\partial U}\Bigr|_G \Bigr) \\
        &{}= \frac{1}{\beta} \sum_{\alpha,ef} \, \chi_{0,abef}^{k\mathbf{q}=0} \, g_{\alpha, ef}^{k\mathbf{q}=0} \, \sigma_{\alpha}^{\mathbf{q}=0},
    \end{aligned}
\end{equation}
where we defined $\sigma_{\alpha}^{\mathbf{q}=0} := \Bigl( \sum_{k'',gh} \tilde{V}_{k''\mathbf{q}=0, gh}^{\alpha} \Bigr)^{-1} \, \sum_{k',cd} \tilde{V}_{k'\mathbf{q}=0,cd}^{\alpha}\,\Bigl( \frac{-\partial \Sigma_{k',cd}}{\partial U}\Bigr|_G \Bigr)$.

With Eqs.~\eqref{equ:dg_du_eph} and \eqref{equ:dg_du_eph_mo}, we derived an \emph{exact} connection, generally valid also beyond DMFT, between the renormalized electron-phonon coupling and the thermodynamic stability criteria of the (multiorbital) Hubbard model. In particular, we show that the vicinity to a Mott MIT, associated to a diverging $dD/ dU$, will, except for peculiar high-symmetry cases, necessarily lead to a divergence of $dn / d\mu$ \emph{as well as} of the renormalized electron-phonon coupling \emph{independent} of the dimensionality of the system.


\clearpage 
\onecolumngrid 
\setcounter{page}{1}
\setcounter{secnumdepth}{1}

\begin{center}
    \textbf{\large Supplemental material: Effective enhancement of the electron-phonon coupling \\ driven by  nonperturbative electronic density fluctuations}\\[0.5em]

    \vspace{1em}
    
    E. Moghadas$^{a}$ \orcidlink{0009-0001-9441-2123}, M. Reitner$^{a}$ \orcidlink{0000-0002-2529-0847}, T. Wehling$^{b}$ \orcidlink{0000-0002-5579-2231}, G. Sangiovanni$^{c}$ \orcidlink{0000-0003-2218-2901}, S. Ciuchi$^{d}$ \orcidlink{0000-0003-1251-1484}, and A. Toschi$^{a}$ \orcidlink{0000-0001-5669-3377}

    \small
    \textit{$^a$Institute of Solid State Physics, TU Wien, 1040 Vienna, Austria} \\
    \textit{$^b$ I. Institute for Theoretical Physics, Universität Hamburg, Notkestraße 9–11, 22607 Hamburg, Germany} \\
    \textit{$^c$Institut für Theoretische Physik und Astrophysik and Würzburg-Dresden Cluster \\ of Excellence ct.qmat, Universität Würzburg, 97074 Würzburg, Germany} \\
    \textit{$^d$Dipartimento di Scienze Fisiche e Chimiche, Università dell’Aquila, Coppito-L’Aquila, Italy}
    
    \vspace{1em}

\end{center}

\thispagestyle{empty}

\twocolumngrid

In this supplemental material we provide details on the model and methods (Sec.~\ref{sec:comp}) as well as on the two-particle formalism and the renormalized el-ph vertex analyzed in the main text (Sec.~\ref{sec:2p}). Subsequently, in Sec.~\ref{sec:hif_exp} we present the high-frequency expansion of $\mathcal{T}_{\mathbf{q}}$, on the basis of which, in Sec.~\ref{sec:ev_prop},  we discuss the spectral properties of the charge response and the renormalized coupling in more detail. In Sec.~\ref{sec:oz}, we explicitly justify the OZ-parametrization of the el-ph vertex. Further, we comment on the validity of our formalism also in the more general case of a next-nearest-neighbor hopping in the Hubbard model in Sec~\ref{sec:tp}. Lastly, in Sec.~\ref{sec:dga}, we discuss a complementary approach to obtain all lowest order el-ph selfenergy diagrams based on the ab-initio dynamical vertex approximation.

\setcounter{equation}{0}
\setcounter{figure}{0}
\setcounter{table}{0}
\setcounter{section}{0}
\renewcommand{\theequation}{S\arabic{equation}}
\renewcommand{\thefigure}{S\arabic{figure}}
\renewcommand{\thetable}{S\arabic{table}}
\renewcommand{\thesection}{S\arabic{section}}

\makeatletter
\renewcommand\section{\@startsection
  {section}
  {1}
  {\z@}
  {-5ex \@plus -1ex \@minus -.2ex}
  {1.2ex \@plus 0.3ex}
  {\normalfont\small\bfseries\centering}
}%
\makeatother

\section{Model and Computational details}\label{sec:comp}
In this work we consider the single-band Hubbard model on a two-dimensional square lattice out of half-filling. The corresponding Hamiltonian reads
\begin{equation}\label{equ:hubbard_H}
    \mathcal{H} = -t\sum_{\braket{i,j}, \sigma}c_{i\sigma}^{\dagger} c_{j\sigma}^{\pdg} + U \sum_i n_{i\uparrow} n_{i\downarrow} - \mu \sum_i (n_{i\uparrow} + n_{i\downarrow}) ,
\end{equation}
where $\braket{...}$ refers to summation between neighboring sites.  $c_{i\sigma}^{\dagger}/c_{i\sigma}$ are the fermionic creation and annihilation operators for lattice site $i$ and spin $\sigma$ with $n_{i\sigma} = c_{i\sigma}^{\dagger} \, c_{i\sigma}$ denoting the density operator. $U$ corresponds to the local, repulsive Hubbard interaction and $\mu$ is the chemical potential which controls the filling of the system. Throughout this work, we set $t = 1/4$. 

We employ the DMFT algorithm for the treatment of the above Hamiltonian, which requires the solution of a self-consistently determined Anderson impurity model \cite{georges1996}. The data throughout the main text and this supplementary material were obtained from a continuous-time quantum Monte Carlo (QMC) impurity solver in the hybridization expansion as implemented in the \textit{w2dynamics} \cite{w2dyn} code-package. 

For all QMC measurements the tight binding dispersion for the square lattice was discretized on a $48\times48$ momentum-grid. After the convergence of the DMFT algorithm the one-particle quantities were recomputed with a single-shot high-statistics run with $\mathcal{O}(10^7)$ measurements. This was used as starting input for the measurement of the two-particle quantities, which were obtained on a frequency grid comprised of $N_{\nu}=130$ positive fermionic and $N_{\omega}=20$ positive bosonic frequencies using $\mathcal{O}(10^6)$ measurements. At the rather high temperatures studied here no asymptotic treatment of the bosonic frequency summations was required while for the fermionic ones the missing high-frequency contribution was approximated with the analytically known $\mathcal{O}(1/\nu^2)$-tail of the bubble term (which is described in more detail in \cite{reitnerDA, moghadasDA}).

For internal momentum integrals in the momentum dependent Bethe-Salpeter equation (BSE), required to obtain the lattice response function, the dispersion was discretized on a $48\times48$ $\mathbf{k}$-grid. For the momentum summations of the static contributions to the self-energy and pairing diagrams of Fig.~\ref{fig:pairing} of the main text we used a momentum grid with $96\times96$ $\mathbf{q}$-points to ensure a converged momentum resolution close to the phase-separation instability where the correlation length becomes particularly large. In contrast, for the dynamic part, which shows no strong momentum dependence, we found a $10\times10$ grid to be sufficient.
\vspace{-5mm}
\subsection{Data Availability}
A data set containing all numerical data and plot scripts used to generate the figures of this publication is publicly available at \cite{data-repo}

\section{Two-particle formalism}\label{sec:2p}

In the main text we investigate the renormalization of the electron-phonon vertex induced by charge fluctuations, since Holstein phonons couple to the electronic density. The central quantity analyzed in the course of this work will be, then, the generalized susceptibility in the \emph{particle-hole} (ph) notation, which for SU(2) symmetric and time-translational invariant systems reads in momentum space 
\cite{rohringer2014phd, rohringer2012}
\begin{equation}\label{equ:chi_gen}
    \begin{aligned}
        {}& \chi_{\sigma \sigma', \mathbf{k}\mathbf{k'}\mathbf{q}}^{\nu \nu' \omega} = \int_0^{\beta} d\tau_1 \, d\tau_2 \, d\tau_3 \, e^{-i \nu \tau_1} \, e^{i (\nu+\omega) \, \tau_2} e^{-i (\nu'+\omega) \tau_3} \\ 
        {}& \times \Bigl\{ \braket{ T_{\tau} \, c_{\sigma, \mathbf{k}}^\dagger(\tau_1) \, c_{\sigma, \mathbf{k+q}}^{\pdg}(\tau_2) \, c_{\sigma', \mathbf{k'+q}}^\dagger(\tau_3), \, c^{\pdg}_{\sigma', \mathbf{k'}}(0) } \\ 
        {}& - \delta_{\mathbf{q}0} \braket{T_{\tau} \, c_{\sigma, \mathbf{k}}^\dagger(\tau_1) \, c_{\sigma, \mathbf{k+q}}^{\pdg}(\tau_2)} \braket{T_{\tau} \, c_{\sigma', \mathbf{k'+q}}^\dagger(\tau_3) \, c_{\sigma',\mathbf{k'}}^{\pdg}(0)} \Bigr\}   ,
    \end{aligned}
\end{equation}
where the external brackets define the grand-canonical thermal averaging $\langle \cdots \rangle = \frac{1}{{\cal Z}} \mbox{Tr} \left[ e^{-\beta {\cal \hat{H}}} \cdots \right]$,  ${\cal Z}$ being the corresponding partition function and $\beta$ the inverse temperature of the system. The fermionic and bosonic momenta are denoted by $\mathbf{k}, \mathbf{k'}$ and $\mathbf{q}$ respectively. Further, $\nu = (2n+1)\pi / \beta$ and $\omega = 2n\pi / \beta$ refer to the fermionic and bosonic Matsubara frequencies respectively, with $n \in \mathbb{Z}$ and $T_{\tau}$ denoting  Wick's time-ordering operator in imaginary times $\tau$. 

We recall that a direct link to the physical charge response can be established by summing over the fermionic Matsubara frequencies as well as spin and momentum indices
\begin{equation}\label{equ:chi_phys}
    \chi^{\omega}_{c, \mathbf{q}} = \frac{2}{\beta^2} \sum_{\substack{\mathbf{k}\mathbf{k'} \\ \nu \nu'}} \left(\chi_{\uparrow \uparrow, \mathbf{k}\mathbf{k'}\mathbf{q}}^{\nu \nu' \omega}  + \chi_{\uparrow \downarrow,\mathbf{k}\mathbf{k'}\mathbf{q}} ^{\nu \nu' \omega}\right) .
\end{equation}

In the SU(2) symmetric case, we can relate the momentum dependent physical response in the charge sector to the two-particle irreducible vertex of the charge channel $\Gamma_c$ through the Bethe-Salpeter equation (BSE) \cite{rohringer2012, rohringer2014phd, rohringer2018}, which is depicted diagrammatically in Fig.~\ref{fig:bse}. The first term in the BSE describes the independent propagation of a particle-hole pair, given by the bubble term  $\chi^{\nu \nu' \omega}_{0,\mathbf{kk'q}} = -2\beta G(\nu, \mathbf{k}) G(\nu+\omega, \mathbf{k+q}) \delta_{\nu \nu'} \delta_{\mathbf{k k'}}$. The remaining terms denote the so-called vertex corrections and correspond to the resummation of infinitely repeated insertions of the irreducible charge vertex $\Gamma_c$ into the bubble term. The aforementioned vertex corrections can also be defined via the full vertex $F_{c, \, \mathbf{k k_q}}^{\nu \nu \omega}$, containing all two-particle scattering events, leading to the following expression for the generalized susceptibility
\begin{equation}
    \chi_{c, \, \mathbf{k k' q}}^{\nu \nu' \omega} = \chi_{0, \, \mathbf{k k' q}}^{\nu \nu' \omega}  - \frac{1}{\beta^2} \sum_{\substack{\mathbf{k_1k_2} \\ \nu_1 \nu_2}} \chi_{0, \, \mathbf{k k_1 q}}^{\nu \nu_1 \omega} F_{c, \, \mathbf{k_1 k_2 q}}^{\nu_1 \nu_2 \omega} \chi_{0, \, \mathbf{k_2 k' q}}^{\nu_2 \nu' \omega}.
\end{equation}
The BSE in its full momentum-dependence can be formulated for the generalized charge susceptibility via \cite{rohringer2012, rohringer2014phd, rohringer2018} 
\begin{equation}\label{equ:bse_c}
    \chi_{c, \, \mathbf{k k' q}}^{\nu \nu' \omega} = \chi_{0, \, \mathbf{k k' q}}^{\nu \nu' \omega}  - \frac{1}{\beta^2} \sum_{\substack{\mathbf{k_1k_2} \\ \nu_1 \nu_2}} \chi_{0, \, \mathbf{k k_1 q}}^{\nu \nu_1 \omega} \Gamma_{c, \, \mathbf{k_1 k_2 q}}^{\nu_1 \nu_2 \omega} \chi_{c, \, \mathbf{k_2 k' q}}^{\nu_2 \nu' \omega} ,
\end{equation}
which can be solved for $\chi_{c, \, \mathbf{k k' q}}^{\nu \nu' \omega}$ 
\begin{equation}\label{equ:bse_inv}
    \chi_{c, \, \mathbf{kk'q}}^{\nu \nu' \omega} = \left[ \left[\chi_{0,\mathbf{kq}}^{\nu \omega}\right]^{-1} \, \delta_{\nu \nu'} \delta_{\mathbf{k k'}} \, + \, \frac{1}{\beta^2} \, \Gamma_{c, \mathbf{kk'q}}^{\nu \nu' \omega} \right]^{-1} ,
\end{equation}
where $\chi_{0,\mathbf{k k' q}}^{\nu \nu' \omega} = \chi_{0,\mathbf{kq}}^{\nu \omega} \delta_{\nu \nu'} \delta_{\mathbf{k k'}}$.

The subsequent sum over the fermionic frequencies and momenta yields the following BSE-formulation of the physical response function
\begin{equation}\label{equ:bse_inv_phys}
    \chi_{c,\mathbf{q}}^{\omega} = \sum_{\substack{\mathbf{k k'} \\ \nu \nu'}} \left[\left[\chi_{0,\mathbf{kq}}^{\nu \omega}\right]^{-1} \, \delta_{\nu \nu'} \delta_{\mathbf{k k'}} \, + \, \frac{1}{\beta^2} \, \Gamma_{c, \mathbf{kk'q}}^{\nu \nu' \omega} \right]^{-1} .
\end{equation}

The inclusion of the correlation effects, associated to this DMFT charge response on the electron-phonon scattering processes yields a renormalization of the el-ph coupling. In particular, the renormalized el-ph vertex, if one neglects all vertex corrections of purely phonon processes, can be obtained from Eq.~\eqref{equ:bse_inv_phys} via:
\begin{equation}
    \tilde{g}_{\mathbf{kq}}^{\nu\omega} = \sum_{\mathbf{k'} \nu'} \left[\chi_{0,\mathbf{kq}}^{\nu \omega}\right]^{-1} \chi_{c, \, \mathbf{k k'q}}^{\nu \nu'\omega} \, g_{0,\mathbf{k'q}}
\end{equation}
where $g^0_{\mathbf{kq}}$ is now a general bare coupling, which can be dependent on momentum. In the here considered case of local Holstein phonons, however, one has $g^0_{\mathbf{kq}}=g^0$. Furthermore, within the DMFT approximation \cite{georges1996}, where only the local part of the irreducible vertex is considered \footnote{Although the irreducible vertex function $\Gamma$ in a selected channel is in general not a local quantity, we recall that in infinite dimensions, where DMFT is exact, the restriction of considering its purely on-site part (i.e., $\Gamma^{\nu \nu' \omega}$) does not imply any error when calculating a BSE-like expression such as Eq.~(\ref{equ:bse_inv}), since any corresponding non-local contribution gets washed out by the internal momentum summations over the $\infty$-dimensional Brillouin zone \cite{georges1996,delre2021}. Of course, this exact result does not apply in finite dimensions, where DMFT becomes an approximation and where, in principle, different choices for the explicit expression of $\Gamma$ in Eq.~(\ref{equ:bse_inv}) would be possible \cite{georges1996}. The most common one is to assume, nonetheless, $\Gamma$ to be purely local, as this approximation (i) allows to minimize the numerical effort and (ii) preserve \cite{nourafkan2019,krien2017,georges1996} the Ward-identities of DMFT, enabling a conserving description\cite{baym61, bickers91}.}, the above expression simplifies to
\begin{equation}\label{equ:eph_hh}
    \begin{aligned}
        \gamma^{\nu \omega}_{\mathbf{q}} &= \tilde{g}^{\nu \omega}_{\mathbf{q}}/g_0 = \left[\chi^{\nu \omega}_{0, \mathbf{q}}\right]^{-1} \sum_{\nu'}\chi^{c, \nu \nu' \omega}_{\mathbf{q}} \\
        &= \sum_{\nu'} \left[\delta_{\nu \nu'} + \frac{1}{\beta^2} \, \Gamma_c^{\nu \nu' \omega} \, \chi_{0, \, \mathbf{q}}^{\nu' \omega}  \right]^{-1}
    \end{aligned}
\end{equation}
where we defined the dimensionless quantity $\gamma$ as the ratio between the renormalized and the bare coupling.

\begin{figure}
    \centering
    \includegraphics[width=0.48\textwidth]{figures/bse.pdf}
    \caption{Diagrammatic representation of the Bethe-Salpeter equation. Here, $\chi_c$ denotes the full charge susceptibility and the first term, $\chi_0$, corresponds to the bubble diagram. $\Gamma_c$ contains all two-particle irreducible diagrams (here in the charge channel within the particle-hole convention).}
    \label{fig:bse}
\end{figure}

\subsection{Transversal ladder-diagrams}
We want to note, that a different choice can be made for the renormalization of the el-ph vertex based on a DMFT calculation. However, as we will discuss in the following, the choice we made captures the leading order corrections close to phase-separation instabilities.

For instance one can start in general from the parquet decomposition \cite{rohringer2012, rohringer2014phd, rohringer2018}. In this case one would define a full vertex $F_{c,\mathbf{kk'q}}^{ladder,\nu\nu'\omega}$ in the charge channel similar to the one in the ladder-D$\Gamma$A approximation \cite{rohringer2018}. In this approach non-local correlations are included in the full vertex via ladder-diagrams in the longitudinal and transversal \textit{ph} channels while the fully irreducible and pairing diagrams are assumed to be local. The full vertex constructed in this approximation can be then used to renormalize the el-ph vertex via
\begin{equation}
    \gamma^{\nu\omega}_{\mathbf{kq}} = \tilde{g}^{\nu\omega}_{\mathbf{kq}}/g_0  = \left[ \mathbb{1} + \frac{1}{\beta^2}\sum_{\nu', \, \mathbf{k'}}F_{c,\mathbf{kk'q}}^{ladder,\nu\nu'\omega} \, \chi^{\nu'\omega}_{0,\mathbf{k'q}} \right]
\end{equation}
In contrast to Eq.~\eqref{equ:eph_hh}, the transversal channel, which transfers a momentum $\mathbf{k'-k}$ through the ladders, will lead to an explicit $\mathbf{k}$-dependence of the renormalized coupling. This contribution originating from the transversal diagrams, which we denote as $\gamma_{\bot}$ can be more formally written as
\begin{equation}
\begin{aligned}
    \gamma^{\nu\omega}_{\bot, \mathbf{kq}} {}&\sim \sum_{\nu'}\int d^2k' \, F_{c,\mathbf{k'-k}}^{\nu \, (\nu+\omega) \, (\nu'-\nu)} \, \chi^{\nu'\omega}_{0,\mathbf{k'q}} \\
    {}&\cong \int d^2k' \, \frac{F_{c,\mathbf{k'-k}=0}^{\nu \, (\nu+\omega) \, (0)}\,\chi^{\nu\omega}_{0,\mathbf{k'q}}}{1 + ((\mathbf{k'-k})\,\xi)^2} \,  + \sum_{\nu'-\nu \, \neq 0}\mathcal{R}(\nu'-\nu)
\end{aligned}
\end{equation}    
where in the last line we have replaced the full vertex with its OZ parametrized form and decomposed it into its static and dynamic components in close analogy to the main text. Furthermore, using the fact that in the strong-coupling regime the bubble term will be only weakly momentum dependent, the above momentum integral will grow only logarithmically with the correlation length $\xi$ upon approaching the phase-transition. Thus, due to the internal summation over the momentum $\mathbf{k'}$ of the transversal diagrams, this contribution will only yield a subleading correction with respect to the longitudinal component, corresponding to Eq.~\eqref{equ:eph_hh}, which scales like $\sim\xi^2$.

\section{High-frequency Expansion\label{sec:hif_exp} of {$\mathcal{T}_{\mathbf{q}}$}}

\begin{figure}
    \centering
    \includegraphics[width=0.45\textwidth]{figures/bubble_diff.pdf}
    \caption{Real and imaginary part of $\beta\mathcal{T}_{\mathbf{q}=0}(\nu)$ as a function of fermionic Matsubara frequencies $\nu$ for increasing values of $U$.}
    \label{fig:bubble_diff}
\end{figure}

\begin{figure*}
    \centering
    \includegraphics[width=\textwidth]{figures/bubble_diff_phase.pdf}
    \caption{Difference of the inverse bubble terms, $\beta \, \mathcal{T}_{\mathbf{q}}(\nu)$, as a function of the fermionic Matsubara frequencies for various values of $U$ and $\mathbf{q}$.}
    \label{fig:bubble_dif_phase}
\end{figure*}

In this section we justify the expansion of the difference between the inverse bubble terms given by
\begin{equation}
    \beta^2 \left( [\chi^{0, \nu \nu'}_{\mathbf{q}}]^{-1} - [\chi_{loc}^{0, \nu \nu'}]^{-1} \right) 
\end{equation}
which encodes the entire momentum dependence of the lattice charge susceptibility in Eq.~\eqref{equ:kappa_bethe} of the main text. 

In the notable case of the infinite-dimensional Bethe lattice the expression for the inverse bubble difference for the uniform and the staggered susceptibility (formally corresponding to the special points $\mathbf{q}=(0,0,0,...)$ and $\mathbf{q}=(\pi, \pi, \pi,...)$ in the infinite-dimensional hyper-cubic) system becomes particularly simple. For instance, for the uniform case one has
\begin{equation}\label{equ:bub_dif_bethe}
[\chi^{0, \nu \nu'}_{\mathbf{q}=0}]^{-1} - [\chi_{loc}^{0, \nu \nu'}]^{-1} = \frac{t^2}{2\beta} \,\delta_{\nu \nu'}     
\end{equation}
with $t$ being the amplitude of the n.n. hopping. For the staggered case, one should  include an additional minus sign in front of $t^2$ \cite{delre2021}. In the case of the two-dimensional square lattice studied here Eq.~\eqref{equ:bub_dif_bethe} can be used, as shown in Appendix A, as an approximation by replacing $t^2$ with an effective $\mathcal{T}_{\mathbf{q}}$ term, corresponding to
\begin{equation}\label{equ:teff_bub}
    \beta^2 \left( [\chi^{0, \nu \nu'}_{\mathbf{q}}]^{-1} - [\chi_{loc}^{0, \nu \nu'}]^{-1} \right) = \beta \mathcal{T}_{\mathbf{q}}(\nu) \delta_{\nu\nu'} \approx  \beta \mathcal{T}_{\mathbf{q}}\delta_{\nu\nu'}.
\end{equation}
Here, a real-valued $\mathcal{T}_{\mathbf{q}}$, slowly varying in the fermionic frequency space, would be required to justify employing Eq.~\eqref{equ:kappa_bethe} for the quantitative analysis of the compressibility in terms of the eigenvalue structure of the local charge susceptibility. In Fig.~\ref{fig:bubble_diff} we plot the real and imaginary parts of $\mathcal{T}_{\mathbf{q}=0}(\nu)$ as a function of the fermionic frequencies for various values of $U$. We indeed observe that the imaginary part is by almost two orders of magnitude smaller than the real component and can be safely neglected. In addition, the real part exhibits a weak $\nu$-dependence for sufficiently large interactions, suggesting that the approximation of Eq.~\eqref{equ:teff_bub} is valid especially in the intermediate-to-strong coupling regime. 

This frequency-feature remains unchanged for finite transfer-momenta. More specifically, we analyzed the $\mathbf{q}$-dependence of this quantity for three different interaction regimes: (i) $U=0$ being the non-interacting limit, (ii) $U=2$ describing the weak coupling, metallic phase and (iii) $U=2.8$ corresponding to the strong coupling, bad-metal case. The results are plotted in Fig~\ref{fig:bubble_dif_phase}, where we show $\beta \, \cal T_{\mathbf{q}}(\nu)$ for varying momenta as a function of the fermionic Matsubara frequencies. We clearly observe a weak momentum dependence for the lower frequencies in the metallic phase in contrast to the bad-metallic one, especially for the limit of $U=0$. More importantly, we can also see that the 
large variation in Matsubara frequencies in the weak-coupling regime vanishes almost entirely for larger interactions. Furthermore, the observation of the equal high-frequency asymptotic in Fig.~\ref{fig:bubble_dif_phase}, independent of the interaction value, leads to the conclusion, that in the strong-coupling regime $\mathcal{T}_{\mathbf{q}}$, as a function of the transfer momentum $\mathbf{q}$, coincides, in general, with the high-frequency expansion, which we illustrate below. 

By extending the results of  Refs.~\cite{georges1996, reitner2020attractive} to the case of finite transfer momenta, we compute the high-frequency asymptotics of the difference between the $\mathbf{q}$-dependent and local bubble susceptibility for zero transfer frequency.

The corresponding bubble term for zero transfer frequency is given by
\begin{equation}\label{equ:bub_q}
    \chi^0_{\mathbf{q}} (\omega=0) = -2\beta \sum_k \frac{1}{z - \epsilon_{\mathbf{k}}} \frac{1}{z - \epsilon_{\mathbf{k+q}}} \quad ,
\end{equation}
where $z = i\nu + \mu + \Sigma(\nu)$ and $\epsilon_{\mathbf{k}}$ denotes the hyper-cubic dispersion relation in $d$ dimensions
\begin{equation}
    \epsilon_{\mathbf{k}} = \frac{-2t}{\sqrt{2 d}} \sum_{i=1}^d \cos (k_i) \quad.
\end{equation}
We can expand the expression in Eq.~\eqref{equ:bub_q} for $|z| \gg \epsilon$, corresponding either to $|\nu|\gg |\epsilon|$, or $|\Sigma(\nu)|\gg |\epsilon|$ for small $\nu$ in the strong coupling regime, to obtain \footnote{The term $\sim \mathcal{O}(z^{-3})$ is linear in $\epsilon_{\mathbf{k}}$ and vanishes after the $\mathbf{k}$-summation.}
\begin{equation}\label{equ:bub_exp}
\begin{aligned}
    {}& \chi^0_{\mathbf{q}} \\
    {}& \cong - 2 \beta \sum_{\mathbf{k}} \left\{ \frac{1}{z^2} + \frac{1}{z^4} \left[ \epsilon_{\mathbf{k}}^2 + \epsilon_{\mathbf{k}} \epsilon_{\mathbf{k+q}} + \epsilon_{\mathbf{k+q}}^2 \right] \right\} + \mathcal{O}(z^{-6})
\end{aligned}
\end{equation}
Using a trigonometric identity to rewrite $\epsilon_{\mathbf{k+q}}$ as 
\begin{equation}
    \epsilon_{\mathbf{k+q}} = \frac{-2t}{\sqrt{2 d}} \sum_{i=1}^d \left( \cos(k_i) \cos(q_i) - \sin(k_i) \sin(q_i) \right)
\end{equation}
and subsequently neglecting all terms linear in $\sin(k_i)$ and $\cos(k_i)$, as those vanish in the momentum integration over $\mathbf{k}$, we can rewrite Eq.~\eqref{equ:bub_exp} as
\begin{equation}
\begin{aligned}
    {}& \chi_{\mathbf{q}}^0 \\
    {}& \cong -2 \beta \sum_{\mathbf{k}} \Biggl\{ \frac{1}{z^2} + \frac{1}{z^4} \Biggr[\epsilon_{\mathbf{k}}^2 + \frac{2t^2}{d} \Biggl( \sum_i \cos^2(k_i) \cos(q_i) \\ 
    {}& + \sum_i \cos^2(k_i) \cos^2(q_i) + \sum_i \left(1-\cos^2(k_i) \right) \sin^2(q_i) \Biggr) \Biggr] \Biggr\}
\end{aligned}
\end{equation}
For the integration over all momenta $k$ we will use the fact that in general
\begin{equation}
    \sum_{\mathbf{k}} \sum_i \cos^2(k_i) f \rightarrow \sum_i f \int \frac{d^d k}{(2 \pi)^d} \cos^2(k_i) = \frac{1}{2} \sum_i f \quad ,
\end{equation}
where $f$ is a function independent of $k_i$. We then obtain the final expression for the $\mathbf{q}$-dependent bubble susceptibility
\begin{equation}
    \chi_{\mathbf{q}}^0 \cong -2\beta \left\{ \frac{1}{z^2} + \frac{1}{z^4}\left[ 2t^2 + \frac{t^2}{d}\sum_{i=1}^d\cos(q_i) \right] \right\} \quad .
\end{equation}

To compute the local bubble susceptibility we have to integrate Eq.~\eqref{equ:bub_q} over all values of $\mathbf{q}$. Then the sums over the two momenta will factorize and we obtain the following expression for the local bubble susceptibility
\begin{equation}
    \chi_{loc}^0 = -2 \beta \left[ \int d\epsilon D(\epsilon) \frac{1}{z - \epsilon} \right]^2 \quad ,
\end{equation}
where the sum over ${\mathbf{k}}$ has been replaced by the integral with the non-interacting density of states $D(\epsilon)$. As we did before, we can expand this expression for large $z$ which results in
\begin{equation}
    \begin{aligned}
        \chi_{loc}^0 \cong{}& -2 \beta \left[ \int d\epsilon D(\epsilon) \left( \frac{1}{z} + \frac{\epsilon^2}{z^3} + \mathcal{O}(z^{-5}) \right) \right]^2 \\
        ={}& -2\beta \left[ \frac{1}{z^2} + \frac{2t^2}{z^4} \right] + \mathcal{O}(z^{-6}) \quad .
    \end{aligned}
\end{equation}
Eventually, it is straightforward to show that the difference between these two terms in the limit of $|z| \gg \epsilon$ will lead to
\begin{equation}\label{equ:bub_dif_d}
    \left[ \chi^0_{\mathbf{q}} \right]^{-1} - \left[ \chi^0_{loc} \right]^{-1} \cong \frac{t^2}{2\beta}\frac{1}{d}\sum_{i=1}^d \cos(q_i) \quad .
\end{equation}
We note, that the hopping parameter here has a different normalization factor than the dispersion relation chosen in the main text for the square lattice, due to the additional $\frac{1}{\sqrt{2 d}}$ factor. Taking this prefactor into consideration, the derived expression for the difference of the inverse of the bubble terms for $d=2$ used in Eq.~\eqref{equ:teff_bub} will yield the following compact expression for $\mathcal{T}_{\mathbf{q}}$ which was also employed in the main text and Appendix A
\begin{equation}\label{equ:teff_hi_freq}
     \beta \mathcal{T}_{\mathbf{q}} \approx \beta \, t^2\Bigl[ \cos(q_x) + \cos(q_y) \Bigr].
\end{equation}

Eqs. ~\eqref{equ:bub_dif_d} and ~\eqref{equ:teff_hi_freq} evidently explain the momentum dependence observed in the numerical results for the strong-coupling regime discussed in the main text. Furthermore they provide an analytical description for the high-frequency behavior of the difference between the inverse of the bubble terms as a function of the transfer momentum and thus explains quantitatively how the sign flip of $\mathcal{T}_{\mathbf{q}}$ occurs between $\mathbf{q}=(0,0)$ and $\mathbf{q}=(\pi,\pi)$.

\section{Spectral properties of the charge response}\label{sec:ev_prop}
\begin{figure}[h!]
    \centering
    \includegraphics[width=0.48\textwidth]{figures/lambda_min.pdf}
    \caption{Lowest real eigenvalues $\lambda_{I}, \, \lambda_{II}$ of the local charge susceptibility $\chi_{loc}^c$ with their respective spectral weights $w_I, \, w_{II}$ and the variation of $-\left[\beta \mathcal{T}_{\mathbf{q}=0}(\nu)\right]^{-1}$ for all fermionic frequencies $\nu$ (blue shaded area) as well as the numerical estimation of $-1/(\beta \mathcal{T}_{eff})$ (blue dashed line). The inset in the lower panel shows the evolution of the spectral weights associated to the lowest eigenvalue $w_I$ and to the effective electron-phonon coupling $\tilde{w}^{\nu=\pi T}_I$ (see Eq.~\eqref{equ:eph_lambda_app} of Appendix B), with and without the contribution of the bubble term as a function of the interaction $U$. The latter is scaled with $\beta$-factors accordingly to ensure all quantities are dimensionless.}
    \label{fig:lambda_min}
\end{figure}

\begin{figure}
    \centering
    \includegraphics[width=\linewidth]{figures/VI_U.pdf}
    \caption{Evolution of the real imaginary part of the first eigenvector with the interaction $U$ in the orthonormal basis, where $V^T V = \mathbbm{1}$.}
    \label{fig:v1_U}
\end{figure}

\begin{figure}
    \centering
    \includegraphics[width=\linewidth]{figures/chi_gamma_small_q.pdf}
    \caption{Upper panel: Static charge susceptibility as a function of $U$ for different, finite momenta $\mathbf{q}$. Lower panel: Same graphs as for the above panel but for the renormalized electron-phonon vertex.}
    \label{fig:chi_vs_gamma}
\end{figure}

In order to formally explain the evolution of the charge susceptibility and the el-ph vertex with increasing interaction strength, as shown in the main text, we will analyze the behavior of the eigenvalue structure of the charge response in the eigenbasis of the local susceptibility. 

In Fig.~\ref{fig:lambda_min} the two smallest, real eigenvalues $\lambda_i$ of the local charge susceptibility $\chi^c_{loc}$ as well as their associated spectral weights $w_i$ as a function of the interaction $U$ are presented. Here, the range of the possible values for the inverse of Eq.~\eqref{equ:teff_bub} at $\mathbf{q}=0$ obtained for different frequencies $\nu$ has been denoted by the blue shaded area. We can immediately note that the variation for $\mathcal{T}_{\mathbf{q}=0}(\nu)$ is relatively small in the bad-metallic region, whereas for the metallic case it becomes significantly larger. This indicates, that for lower values of $U$, the Bethe lattice approximation for $\chi_{\mathbf{q}=0}$ of the square lattice is not well justified for weak coupling but indeed becomes a good approximation in the bad metallic regime of our interest.

The blue-dashed line marks instead a numerical estimate of $-1 / \beta \mathcal{T}_{\mathbf{q}=0}$ which we denote, here, as $-1/\beta\,\mathcal{T}_{eff}$ \footnote{This is estimated by varying the lowest eigenvalue until the susceptibility diverges, which we can regard as the \emph{actual} value of $\mathcal{T}_{eff}$, for details see \cite{reitner2020attractive} Appendix I.}. As discussed in Ref.~\cite{reitner2020attractive}, since the smallest eigenvalue $\lambda_I<0$ has a negative weight $w_I<0$, it will contribute to an overall enhancement of the charge response, whereas the second smallest eigenvalue $\lambda_{II}$, which although being negative as well, has a positive spectral weight $w_{II}>0$, will lead to an overall suppression of the susceptibility. We immediately observe that around $U=2.4$ the difference between $\lambda_I$ and $\mathcal{T}_{eff}$ becomes vanishingly small explaining the large enhancement of the static, uniform charge susceptibility. Another important aspect worth mentioning, is the fact that the spectral weight of the first eigenvalue is essentially zero in the metallic regime, which strongly suppresses the contribution of $\lambda_I$ to the compressibility. Only for $U\gtrsim2.4$, it yields a sizable value, which explains, to some degree, the \emph{asymmetric shape} of the compressibility and also the electron-phonon coupling in Figs.~\ref{fig:chi_q_U} and ~\ref{fig:gamma_q_U} of the main text: Only in the stronger correlated regime, where the spectral weight of the lowest eigenvalue becomes sizable, a significant enhancement of the charge response or the electron-phonon coupling for lower transfer momenta can be observed. 

The origin of this behavior can be traced to the different impact of the particle-hole symmetry violation for the same value of electronic density in the two regimes, which is directly reflected in the structure of the first eigenvector $V_I(\nu)$ presented in Fig.~\ref{fig:v1_U}. In this specific case we represent the eigenvector as a function of fermionic frequencies for varying $U$ values in an orthonormal basis $V^T \, V =  \mathbbm{1}$, where it can be shown that either the real part of the vector is symmetric and the imaginary part is antisymmetric or vice verse \cite{reitner2023nh}. We note, that the real part is entirely antisymmetric and hence cannot contribute to the spectral weight, as in that case $\sum_{\nu}\,V_I \approx 0$. The frequency sum over the imaginary part on the other hand yields a small negative contribution to the spectral weight $w_{\alpha}$ in the metallic phase, as a consequence of the small particle-hole symmetry violation, and only acquires a larger magnitude with increasing interaction, which explains the behavior of the spectral weight observed in Fig.~\ref{fig:lambda_min}.

At this point it is also important to discuss the evident difference in the overall magnitude between the charge susceptibility and the renormalized vertex. Evidently, this difference, which has been also discussed in the main text, can be ascribed to the different spectral weights in Eqs.~\eqref{equ:gen_susc_app} and ~\eqref{equ:eph_lambda_app} of Appendices A and B respectively. In the case of the renormalized coupling the weight $\tilde{w}_{\alpha}^{\nu}$ contains the inverse of the bubble susceptibility. This quantity can become quite large for stronger interaction values and result in an overall enhancement of the weight compared to $w_I$. In the inset of the lower panel of Fig.~\ref{fig:lambda_min} we compare the absolute value of the spectral weights $w_I$ with $\tilde{w}_{\alpha}^{\nu}$, once with (bold line and full markers) and once without the contribution of the inverse bubble term (dashed line with empty markers) on a logarithmic scale each scaled accordingly with $\beta$ factors to ensure the same dimensionality in all quantities. We observe that only with the inclusion of the bubble term the $\tilde{w}_{\alpha}^{\nu}$ will become significantly larger than $w_I$, demonstrating that the difference in magnitude between the charge response and the renormalized electron-phonon coupling originates from the inverse bubble term. 

This specific relationship between the charge response and the electron-phonon vertex enhancement might have also caused some of the contradicting interpretations in the past literature. In particular, in Ref.~\cite{arrigoni2003} the increase of the el-ph coupling was not attributed to the proximity to a phase-separation instability as the corresponding charge response was monotonically decreasing with the Hubabrd $U$. However, due to the additional factor coming from the inverse bubble term an increase of the el-ph vertex can indeed become visible at larger momenta than those for which the charge response shows any enhancements. An example is shown in Fig.~\ref{fig:chi_vs_gamma} where we compare the charge response and the renormalized vertex for various finite $\mathbf{q}$ as a function of the interaction. In the upper panel, an enhancement of the charge susceptibility for $\mathbf{q}=(\pi/8, \pi/8)$ is observed, while for smaller momenta the response decreases monotonically. However, the electron-phonon vertex for the same transfer-momenta, shown in the lower panel, exhibits an upturn for large enough values of $U$. This indicates, that despite being in the vicinity of a phase-transition, a significant enhancement of the el-ph vertex can be obtained without observing a similar trend in the charge susceptibility yet.

\section{correlation length}\label{sec:oz}
\begin{figure}
    \centering
    \includegraphics[width=0.45\textwidth]{figures/gamma_q_oz.pdf}
    \caption{Renormalized electron-phonon vertex as a function of momentum $\mathbf{q}$ along the Brillouin zone diagonal for the first eight fermionic Matsubara frequencies. The dashed line denotes an OZ fit to the renormalized vertex, where the same value for the correlation length $\xi$ was used, obtained from the momentum dependence of $\gamma_{\mathbf{q}}$ evaluated at $\nu=\pi T$. The inset shows the spectral weight $\tilde{w}_{\alpha}^{\nu}$ as a function of $\mathbf{q}$ for the first eight Matsubara frequencies.}
    \label{fig:gamma_q_oz}
\end{figure}

\begin{figure}[h!]
    \centering
    \includegraphics[width=0.45\textwidth]{figures/corr_len_ev.pdf}
    \caption{Correlation length of the charge response (left panel) and the renormalized electron-phonon coupling (right panel) as a function of $U$ once computed in terms of the eigenvalues of the local susceptibility (red) and once extracted via a fit (green).}
    \label{fig:xi}
\end{figure}

In the main text we assumed an Ornstein Zernike expression not only for the physical charge susceptibility, which is certainly legitimate close to a second order phase transition, but also for the renormalized electron-phonon vertex. For the latter quantity, this assumption appears less obvious. Therefore, in this section we will derive the Ornstein-Zernike (OZ) expressions for lattice charge susceptibility as well as the renormalized coupling in terms of the eigenvalue structure of the local impurity susceptibility and prove that close to the transition the correlation lengths of both quantities are equivalent.

At the critical endpoint, where the isothermal charge susceptibility diverges, marking a second order phase transition, the correlation length $\xi$ will also diverge. In this parameter regime, for small transfer momenta $\mathbf{q}$ the charge response can be parametrized by an Ornstein-Zernike correlation function \cite{goldenfeld, Moriya1985}
\begin{equation}\label{equ:oz_chi}
    \chi_{\mathbf{q}}^c \approx \frac{\chi_{\mathbf{q}=0}^c}{1 + (\xi \mathbf{q})^2}
\end{equation}
In order to reformulate the Ornstein-Zernike expression of the isothermal compressibility in the eigenbasis of the local susceptibility, we first expand the difference of the inverse bubble in terms of small transfer momenta $\mathbf{q}$. In two-dimensions, a small-$\mathbf{q}$ expansion of the cosine in Eqs.~\eqref{equ:bub_dif_d} and ~\eqref{equ:teff_hi_freq} will read
\begin{equation}
    \left[ \chi^0_{\mathbf{q}} \right]^{-1} - \left[ \chi^0_{loc} \right]^{-1} \approx \frac{t^2}{\beta}\Bigl[ \cos(q_x) + \cos(q_y) \Bigr] \approx \frac{t^2}{\beta} \Bigl[2-\frac{1}{2} \mathbf{q}^2\Bigr] \, .
\end{equation}
Subsequently, we can rewrite the BSE, Eq.~\eqref{equ:bse_dmft}, in the eigenbasis of the local susceptibility to obtain a modified form of Eq.~\eqref{equ:kappa_bethe}
\begin{equation}
    \chi_{\mathbf{q}}^{c, \nu\nu'} \simeq \beta^2\sum_{\alpha} V_{\alpha}(\nu) \Biggl[ {\lambda}_{\alpha}^{-1} + {2\beta t^2} + \epsilon \mathbf{q}^2 \Biggr]^{-1} V_{\alpha}^{-1}(\nu').
\end{equation}
with $\epsilon=-\beta t^2/2$. Simple algebraic manipulations will give
\begin{equation}\label{equ:gen_chi_alpha}
    \chi_{\mathbf{q}}^{c, \nu\nu'} \simeq \beta^2\sum_{\alpha} V_{\alpha}(\nu) \frac{\chi_{\alpha}}{1 + \epsilon \, \chi_{\alpha} \, \mathbf{q}^2} V_{\alpha}^{-1}(\nu')
\end{equation}
where we have defined $\chi_{\alpha}^{-1} = {\lambda}_{\alpha}^{-1} + 2\beta t^2$. The correlation length associated to one eigenvalue can be extracted from the above expression and corresponds to 
\begin{equation}
    \xi_{\alpha}^2 = \epsilon \, \chi_{\alpha}
\end{equation}
From Eq.~\ref{equ:gen_chi_alpha} we can obtain the physical charge susceptibility by summing over the fermionic degrees of freedom
\begin{equation}\label{equ:chi_oz_alpha}
    \chi_{\mathbf{q}}^c \simeq \sum_{\alpha} \frac{\chi_{\alpha} \, w_{\alpha}}{1 + \xi_{\alpha}^2 \, \mathbf{q}^2}
\end{equation}
with $w_{\alpha} = \sum_{\nu} V_{\alpha}(\nu) \times \sum_{\nu'} V_{\alpha}^{-1}(\nu') $. In close analogy, we can obtain an expression for the renormalized electron-phonon coupling in Eq.~\eqref{equ:eph_hh}
\begin{equation}\label{equ:gamma_oz_alpha}
    \gamma^{\nu}_{\mathbf{q}} = \left[\chi^{0, \nu}_{\mathbf{q}} \right]^{-1} \, \sum_{\nu'} \chi_{\mathbf{q}}^{c, \nu\nu'} \simeq \sum_{\alpha} \frac{\chi_{\alpha} \, \tilde{w}^{\nu}_{\alpha}}{1 + \xi_{\alpha}^2 \, \mathbf{q}^2}
\end{equation}
with $\tilde{w}^{\nu}_{\alpha} = \beta^2 \, \left[ \chi^{0,\nu}_{\mathbf{q}}\right]^{-1} \, V_{\alpha}(\nu) \, \sum_{\nu'} V_{\alpha}^{-1}(\nu')$. It is worth to recall once again, that in the bad-metallic or strong coupling regime, any momentum structure of the bubble term will be relatively small due to the large local selfenergy in its denominator, leading to an essentially momentum independent spectral weight $\tilde{w}^{\nu}_{\alpha}$. As a consequence, the only $\mathbf{q}$-dependence in the expression of Eq.~\eqref{equ:gamma_oz_alpha} will exclusively originate from the denominator. Hence, the el-ph vertex can be parametrized quite well by an OZ function independent of the fermionic frequency $\nu$. In Fig.~\ref{fig:gamma_q_oz} we show the $\mathbf{q}$-dependence along the Brillouin zone diagonal for the first eight frequencies of the renormalized vertex $\gamma_{\mathbf{q}}^{\nu}$ for the hole-doped Hubbard model at the critical interaction value $U=2.4$. Here we clearly observe the momentum structure reminiscent of an OZ correlation function for the lowest frequencies. This is confirmed by a fit of the OZ expression to the data which is plotted with dashed lines. We want to stress here that for the fits the same value of the correlation length $\xi$, that was obtained from the momentum dependence of $\gamma_{\mathbf{q}}$ at $\nu=\pi T$, was used. For higher frequencies the renormalized vertex becomes vanishingly small and loses the momentum dependence entirely rendering a fit to an OZ expression impossible. This can be understood from our data upon inspecting the frequency structure of the associated spectral weight $\tilde{w}_{\alpha}^{\nu}$. The inset of Fig.~\ref{fig:gamma_q_oz} shows this quantity, which, as expected, is almost constant as a function of momentum, for the lowest (most dominant) eigenvalue as a function of $\mathbf{q}$ for the first eight Matsubara frequencies. Since $\tilde{w}_I^{\nu}$ vanishes essentially for larger Matsubara frequencies ($n>4$), also the corresponding expression for $\gamma_{\mathbf{q}}^{\nu}$ becomes approximately zero. We note that, in any case, already from the expression in Eq.~\eqref{equ:gamma_oz_alpha} one can deduce that the OZ form is valid for all frequencies, as long as the spectral weight remains finite and momentum independent.   

Finally, we can derive an explicit expression for the overall correlation length of the charge response and the electron-phonon coupling in terms of the contributions from the different eigenvalues. First, we expand the fractions in Eq.~\eqref{equ:chi_oz_alpha} and subsequently discard terms of order $\mathcal{O}(\mathbf{q}^3)$
\begin{equation}
    \begin{aligned}
        &{}\chi_{\mathbf{q}}^c \simeq \sum_{\alpha} \frac{\chi_{\alpha} \, w_{\alpha}}{1 + \xi_{\alpha}^2 \, \mathbf{q}^2} \\
        &{} = \frac{\prod_{k\neq I}(1 + \xi_k^2 \mathbf{q}^2)\chi_I w_I + \prod_{k\neq II}(1 + \xi_k^2 \mathbf{q}^2)\chi_{II} w_{II} + ...}{\prod_{\alpha}(1 + \xi_{\alpha}^2 \mathbf{q}^2)} \\
        &{}\approx \frac{(1 + \sum_{k\neq I}\xi_k^2 \mathbf{q}^2)\chi_I w_I + (1 + \sum_{k\neq II}\xi_k^2 \mathbf{q}^2)\chi_{II} w_{II} + ...}{1 + \sum_{\alpha} \xi^2_{\alpha}\mathbf{q}^2} \\
    \end{aligned}
\end{equation}
In the above we can now collect the constant and $\mathbf{q}$-dependent terms to obtain
\begin{equation}
    \begin{aligned}       
        &{}\chi^c_{\mathbf{q}} \approx \frac{\sum_{\alpha} \chi_{\alpha}w_{\alpha} + \mathbf{q}^2 \, \sum_{\alpha} (\sum_{k\neq\alpha} \xi_k^2) \chi_{\alpha} w_{\alpha}}{1 + \sum_{\alpha} \xi^2_{\alpha}\mathbf{q}^2} \\
        &{} \approx \frac{\sum_{\alpha} \chi_{\alpha}w_{\alpha}}{1 + \sum_{\alpha}\xi^2_{\alpha}\mathbf{q}^2} \, \Biggl[1 - \mathbf{q}^2\, \frac{\sum_{\alpha} (\sum_{k\neq\alpha} \xi_k^2) \chi_{\alpha} w_{\alpha}}{\sum_{\alpha} \chi_{\alpha}w_{\alpha}} \Biggr]^{-1}
    \end{aligned}
\end{equation}
where in the last line we have used $1+\epsilon\,\mathbf{q}^2 \approx (1-\epsilon\,\mathbf{q}^2)^{-1}$ for $|\mathbf{q}| \ll 1$. Eventually, we expand the denominator where we again will only keep terms up to second order in $\mathbf{q}$.
\begin{equation}
    \begin{aligned}
        &{} \chi^c_{\mathbf{q}} \\
        &{} \approx \sum_{\alpha} \chi_{\alpha}w_{\alpha} \, \Biggl[1 + \sum_{\alpha} \xi_{\alpha}^2 \mathbf{q}^2 - \mathbf{q}^2\, \frac{\sum_{\alpha} (\sum_{k\neq\alpha} \xi_k^2) \chi_{\alpha} w_{\alpha}}{\sum_{\alpha} \chi_{\alpha}w_{\alpha}} \Biggr]^{-1} \\
        &{} = \sum_{\alpha} \chi_{\alpha}w_{\alpha} \, \Biggl[1 + \left(\sum_{\alpha} \chi_{\alpha} w_{\alpha}\right)^{-1} \sum_{\alpha} \xi_{\alpha}^2 \chi_{\alpha} w_{\alpha} \, \mathbf{q}^2\Biggr]^{-1}
    \end{aligned}
\end{equation}
From the above equation one can extract the overall correlation length, since $\sum_{\alpha}\chi_{\alpha}w_{\alpha} \equiv \chi_{\mathbf{q}=0}$. Hence, the final expression for the correlation length becomes
\begin{equation}\label{equ:xi_alpha}
    \xi^2 \simeq \frac{\sum_{\alpha}\xi^2_{\alpha}\chi_{\alpha}w_{\alpha}}{\sum_{\alpha}\chi_{\alpha}w_{\alpha}}
\end{equation}
The same expression holds for the electron-phonon coupling. In that case, one merely has to replace $w_{\alpha} \to \tilde{w}^{\nu}_{\alpha}$. 

Furthermore, writing the correlation length in this way, highlights that close to the phase instability the sum will be dominated by the contribution of the first eigenvector. Then, for either the charge response or the electron-phonon coupling, the correlation length will be given by
\begin{equation}
    \xi^2 \approx \xi_I^2 + \frac{\sum_{\alpha\neq I}\xi^2_{\alpha}\chi_{\alpha}w_{\alpha}}{\chi_Iw_I}
\end{equation}
where the second term will become negligibly small. Thus, to a first approximation, the correlation length of the charge susceptibility and the renormalized coupling will be equivalent. This is shown in Fig.~\ref{fig:xi}, where we have plotted their respective correlation lengths obtained with Eq.~\eqref{equ:xi_alpha} for both quantities and additionally compared them with the correlation lengths extracted with a fit of Eq.~\eqref{equ:oz_chi} to the numeric data corresponding to the hole-doped Hubbard model presented in the main text. As before, for the electron-phonon coupling we fixed the fermionic Matsubara frequency to the lowest possible value $\nu= \pi T$. We immediately observe a clear-cut agreement of all the plotted quantities, proving, in addition to the fact that the renormalized coupling exhibits an OZ behavior, that close to the phase instability the correlation lengths of the charge susceptibility and the electron-phonon coupling are essentially equivalent. We should also note, at the same time, that for values of $U$ below the instability, i.e. $U<2.4$ the above considerations do not hold, as we are not in a regime where we expect the parametrization with an OZ expression as an appropriate approximation.

\section{Hubbard model with extended hopping}\label{sec:tp}

\begin{figure}[h!]
    \centering
    \includegraphics[width=0.45\textwidth]{figures/tp_eph.pdf}
    \caption{Upper panel: static charge susceptibility $\chi^c_{\mathbf{q}}(\omega=0)$ as a function of $U$ and $\mathbf{q}$ at $\beta=46.75$ and $n=1$ for the Hubbard model with $t'=-0.15t$. Lower panel: Renormalized electron phonon coupling $\gamma_{\mathbf{q}}$ evaluated at the lowest fermionic and bosonic Matsubara frequencies as a function of $U$ and various momenta $\mathbf{q}$.}
    \label{fig:tp_eph}
\end{figure}

\begin{figure}[h!] 
    \centering
    \includegraphics[width=0.45\textwidth]{figures/tp_lambda_min.pdf}
    \caption{Lowest real eigenvalues $\lambda_{I}, \, \lambda_{II}$ of the local charge susceptibility $\chi_{loc}^c$ with their respective spectral weights $w_I, \, w_{II}$ and the variation of $-\left[\beta \mathcal{T}_{\mathbf{q}=0}(\nu)\right]^{-1}$ for all fermionic frequencies $\nu$ (blue shaded area) as well as the numerical estimation of $-1/(\beta \mathcal{T}^2_{eff})$ (blue dashed line). The inset in the lower panel shows the evolution of the spectral weights associated to the lowest eigenvalue $w_I$ and the effective electron-phonon coupling $\tilde{w}^{\nu=\pi T}_I$ ((see Eq.~\eqref{equ:eph_lambda_app} of Appendix B)), once with and without the contribution of the bubble term, with the interaction $U$. The latter is scaled with $\beta$-factors accordingly to ensure all quantities are dimensionless.}
    \label{fig:lambda_min_tp}
\end{figure}

As mentioned in the main text, in order to observe an enhanced charge susceptibility, particle-hole symmetry must be broken. This is possible either via doping or geometric frustration. 

In the following section we will show that the inclusion of the latter, i.e. an additional next-nearest-neighbor (nnn) hopping term $t'$, is indeed sufficient to trigger an enhancement of the charge response even at half-filing.

Similarly to the analysis of the hole-doped Hubbard model, we will study the effect of electronic correlations on the renormalization of the electron-phonon coupling as a function of $U$. Fixing $t'=-0.15t$ locates the Mott-MIT at slightly below $\beta=46.75$, where we observe a non-monotonous behavior of the charge response for interaction values between $U=2.34$ and $U=2.35$. In Fig.~\ref{fig:tp_eph} the results for the static charge susceptibility and renormalized electron-phonon coupling for various transfer momenta along the Brillouin zone diagonal are presented. We immediately observe, in stark contrast to the hole-doped case, that the enhancement of the compressibility, which is shown in the upper panel of Fig.~\ref{fig:tp_eph}, at around $U=2.348$, is relatively small and very sharp (note the $U$-scale). In addition, for all calculated interaction values the $\chi^c_{\mathbf{q}}$ response at $\mathbf{q}=(\pi,\pi)$ is dominating over small transfer momenta (although \textit{exactly} at the critical point, the charge susceptibility at $\mathbf{q}=0$ will eventually diverge). Nonetheless, the renormalized electron-phonon vertex, shown in the panel below, exhibits a very prominent forward scattering peak, which not only is larger by two orders of magnitude than the enhancement of the charge response but also is significantly larger than the large $\mathbf{q}$ region. 

This behavior can be quantitatively understood upon inspecting the spectral decomposition of the charge response or renormalized electron-phonon coupling in the eigenbasis of the auxiliary generalized susceptibility of the auxiliary AIM associated to the DMFT solution. In the upper panel of Fig.~\ref{fig:lambda_min_tp} we present the evolution of the two lowest real eigenvalues $\lambda_I, \lambda_{II}$ of the local charge susceptibility with the interaction $U$ alongside the divergence criterion $-(\beta \mathcal{T}_{eff})^{-1}$ denoted by a blue solid line. Here, the blue shaded area represents again the variation of the inverse of $\beta\mathcal{T}_{\mathbf{q}}(\nu)$ in frequency space. We note, that the divergence criterion is fulfilled close to $U=2.348$ explaining the peak we find for the isothermal compressibility. The lower panel of Fig.~\ref{fig:lambda_min_tp} displays the behavior of the spectral weights of the two lowest eigenvalues $w_I, w_{II}$ as a function of the interaction. Here, we can immediately recognize the cause of the relatively small enhancement of the compressibility, namely the vanishingly small spectral weight $w_I$ of the lowest eigenvalue. In the inset of the lower panel, we show $w_I$ as a function of $U$ on a logarithmic scale to demonstrate that the spectral weight is still finite, proving that the enhancement seen for the compressibility is indeed a result of the lowest eigenvalue fulfilling the divergence criterion of Eq.~\eqref{equ:kappa_bethe}. Additionally, in close analogy to the hole-doped case, we compare the spectral weight with $\tilde{w}_I$, with and without the contribution of the bubble susceptibility. We find here as well, that only with the inclusion of the bubble term, the spectral weight of the renormalized vertex can become significantly larger than $w_I$ thus leading to an overall larger contribution of $\lambda_I$ to the electron-phonon coupling than to the enhancement of the charge response.

It is worth noting the following: We have selected here, for demonstration purposes, the complimentary case with respect to the manuscript of half filling with geometric frustration. However, in the more realistic situation of out-of-half-filling \textit{and} geometric frustration an overall larger enhancement effect on the generalized susceptibility and on the el-ph coupling could be expected.

\section{D$\Gamma$A-like approximation}\label{sec:dga}
Although the diagrams shown in the left panel of Fig.~\ref{fig:pairing} of the main text are important second-order corrections to the electronic selfenergy and pairing interaction, many other diagrams, potentially at even higher orders of the bare coupling can become relevant. 

To illustrate on a more formal level, what is to be expected from this strong interplay between electronic and lattice degrees of freedom close to the Mott transition, we adopt a perturbative resummation of two-particle diagrams inspired by the Abinitio-D$\Gamma$A \cite{toschi2011, julich2011lda+dmft, galler2017} approach. The core assumption of the Abinitio-D$\Gamma$A is to extend the local DMFT irreducible vertex $\Gamma$ by a bare non-local Coulomb interaction. Then, similarly to the D$\Gamma$A approach the full vertex $F$ is constructed from ladders in the $ph$ and $\bar{ph}$-channels. The additional non-local bare Coulumb interaction $V_{\mathbf{q}}$ in the BSE of the $ph$-ladders yields RPA screening diagrams, which when inserted in the SDE will resemble the \textit{GW}-selfenergy. What is however neglected in this approach are nonlocal spatial correlations in $\Gamma$ beyond the bare Coulomb interaction.

In close analogy to the Abinitio-D$\Gamma$A we can obtain all relevant diagrams perturbatively in the bare phonon coupling $g_0$ that DMFT can produce for the Hubbard-Holstein model, by augmenting the irreducible vertex $\Gamma$ of the impurity with a bare phonon-mediated interaction
\begin{equation}\label{equ:bare_dynU}
V_{\sigma\sigma'}^{\nu\nu'\omega} = g_0^2 \, \Bigl( D_{0}^{\omega}-\delta_{\sigma\sigma'}D_{0}^{\nu'-\nu} \Bigr)  
\end{equation}
where $\sigma$ denotes the spin index and $D_0$ is the bare phonon propagator. Defining the phonon-mediated interaction in the charge channel as $V_c = V_{\uparrow\uparrow} + V_{\uparrow\downarrow}$ allows us to write the new irreducible charge vertex, $\tilde{\Gamma}_c$ as
\begin{equation}
    \tilde{\Gamma}_c = \Gamma_c \, + \, g_0^2\Bigl(2\,D_0^{\omega} - D_0^{\nu'-\nu}\Bigr)
\end{equation}

Subsequently, as in conventional ladder-D$\Gamma$A we construct the full vertex $\tilde{F}$ of the Hubbard-Holstein using the BSE in the particle-hole channel. The momentum-dependent self-energy can be obtained via the Schwinger-Dyson equation of motion (EOM) \cite{abrikosov2012methods, rohringer2014phd}, which uses the full vertex as the central component. The charge sector, which in this particular case, incorporates the dominating fluctuations will yield the following contributions to $\tilde{F}$
\begin{equation}    \tilde{F}_{c,\mathbf{q}}^{\nu\nu'\omega} = F_{c,\mathbf{q}}^{\nu\nu'\omega} - \frac{1}{\beta^4}\sum_{\nu_1\nu_2} F_{c,\mathbf{q}}^{\nu\nu_1\omega} \, \chi_{0,\mathbf{q}}^{\nu_1\omega} \, V_c^{\nu_1\nu_2\omega} \, \chi_{0,\mathbf{q}}^{\nu_2\omega} \, \tilde{F}_{c,\mathbf{q}}^{\nu_2\nu'\omega} \end{equation} 
where $F_{c, \mathbf{q}}$ is the purely electronic lattice vertex in the charge channel. This geometric series, entails all possible orders of the interaction $V^{\nu\nu'\omega}$ from Eq.~\eqref{equ:bare_dynU} and can be viewed to some extent as a RPA-like screening of the purely electronic full vertex $F$. In the same spirit of Abinitio-D$\Gamma$A our approach neglects irreducible vertex-corrections beyond the bare phonon-mediated interaction between electrons. This approximation is insofar justified, since for the el-ph selfenergy dependence on the correlation length, the aforementioned phonon-mediated irreducible vertex corrections can be safely neglected as they are purely local for the Hubbard-Holstein model. Thus, the internal momentum integrals will not affect the scaling with respect to $\xi$.

In any case, we can immediately see that a truncation of this series to the lowest order, i.e. $\mathcal{O}(g_0^2)$ (or equivalently $\mathcal{O}(V_c)$) despite an arbitrarily small bare coupling, will be rendered impossible by the divergences of $F_{c, \mathbf{q}}$ in the momentum integrations occurring in the EOM. In fact, we observe a divergence originating from this ladder resummation that scales like $\sim\xi^{2n} \, (\sim\xi^{2n-2}\log\xi)$ for even (odd) orders $n$ of the phonon interaction $V_c$, suggesting a breakdown of this perturbative treatment of the coupled electron-phonon problem at hand.  

\end{document}